\documentclass{JHEP3}
\usepackage{amsmath}
\usepackage{epsfig,comment}
\usepackage{amssymb,amsfonts,nicefrac}

\newcommand{\oao}[2]{{#1\atopwithdelims[]#2}}

\def\zi{\mathbb{Z}}
\def\ci{\mathbb{C}}

\def\di{\text{d}}

\def\slr{\text{SL(2},\mathbb{R})}
\def\slc{\text{SL(2},\mathbb{R})/\text{U}(1)}

\def\p{\partial}
\def\pb{\bar{\partial}}
\def\cn{\mathcal{N}}
\def\er{\mathbb{R}}

\DeclareRobustCommand{\sfrac}[2]{\hspace{0.1em}%
  \raisebox{-0.5ex}{${ #1}$}\hspace{-0.1em} \backslash \hspace{-0.07em}%
  \raisebox{0.5ex}{${ #2}$}}

\title{ \boldmath Non-Critical String Duals of  $\cn = 1$ Quiver Theories 
\unboldmath }

\author{
        Dan Isra\"el\\ 
  Racah Institute of Physics, The Hebrew University\\ 91904 Jerusalem, Israel \\
        E-mail:  \email{israeld@phys.huji.ac.il}
}

\abstract{
We construct $\cn$ =1 non-critical strings in four dimensions dual 
to strongly coupled $\cn$=1 quiver gauge theories in the Coulomb phase, 
generalizing the string duals of Argyres-Douglas superconformal fixed 
points in $\cn$=2 gauge theories. They are the first examples of superstring vacua with an exact worldsheet 
description dual to chiral $\cn$=1 theories. We identify the dual of the 
non-critical superstring using a brane setup describing the field theory in the classical limit. 
We analyze the spectrum of chiral operators in the strongly coupled regime and show 
how worldsheet instanton effects 
give non-perturbative information about the gauge theory. 
We also consider aspects of D-branes relevant for the holographic 
duality.  
} 

\preprint{hep-th/0512166}

\begin{document}
\section{Introduction}
Four dimensional little string theories (\textsc{lst}), or four-dimensional non-critical 
strings~\cite{Ooguri:1995wj,Giveon:1999zm,Pelc:2000hx}, 
are very important examples of holographic dualities in string 
theory. Unlike the usual \textsc{AdS}$_5$/\textsc{cft}$_4$ 
correspondence~\cite{Maldacena:1997re}, 
the string background has no Ramond-Ramond fluxes and is often exactly solvable.\footnote{However, 
some of the most interesting examples, dual to confining $\cn =1$ gauge 
theories~\cite{Maldacena:2000yy}, don't have a known exact worldsheet description.}
They are gravitational duals of non-local theories whose 
infrared dynamics is described by four-dimensional $\mathcal{N}=2$  gauge theories.
In opposition with type \textsc{iib} six-dimensional 
\textsc{lst}~\cite{Seiberg:1997zk,Aharony:1998ub,Giveon:1999px,Ooguri:1995wj}, flowing 
in the infrared to free $\cn = (1,1)$ \textsc{sym}$_6$,  
they probe the gauge theory far from the semi-classical regime, near a strongly coupled 
fixed point. 
These models can be viewed either as coming from NS5-branes wrapped 
on a Seiberg-Witten curve~\cite{Klemm:1996bj}
or equivalently from the decoupling limit of \textsc{cy}$_3$ singularities~\cite{Ooguri:1995wj}. 
In both cases the infrared limit of the dual, non-gravitational theory is an $\cn = 2$ gauge 
theory in a Coulomb phase, and more specifically for the examples that we study pure 
$\cn = 2$ \textsc{sym} theory in the neighborhood of a strongly coupled 
superconformal fixed point, of the Argyres-Douglas type~\cite{Argyres:1995jj}. 
More recently, these non-critical strings were related to $\cn = 1$ \textsc{sym} 
at large large N~\cite{Bertoldi:2005yv}. 

From the string theory point of view, these models can be studied using exact worldsheet 
conformal field theory methods. It is fortunate that 
we have such a worldsheet \textsc{cft} description since the string theory, being non-critical, does not have 
a good supergravity limit, i.e. with gradients of the background fields small 
w.r.t. the string scale.  However, in the regime where the holographic
duality is valid we focus on the neighborhood of a 
superconformal fixed point of the gauge theory, for which 
the coupling constant of the string dual diverges. Then, when the singularity 
is resolved in order for the string coupling to be finite, one cannot have a large hierarchy between 
the mass scales of the effective gauge theory and the little string theory scale while 
keeping the string coupling small, and the gravitational dual will encode little 
string theory physics which is not field-theoretical. Nevertheless the far infrared physics of 
the theory, corresponding to localized excitations in the bulk, will still be 
well described by gauge theory. We argue, in our particular example, that the worldsheet 
instanton effects --~related to the duality between the supersymmetric 
coset $\slc$ and the $\cn = 2$ Liouville theory (called also \textsc{fzz} duality)~\cite{cigdual,Giveon:1999px}~-- 
indeed contain the information about the exact Seiberg-Witten 
low energy effective action the gauge theory, i.e. the 
masses of the \textsc{bps} dyons as functions of the coordinates on the moduli 
space of $\cn = 2$ \textsc{sym} theory.

We will show in this work how to extend this construction to 
$\cn = 1$ models, that are non-critical string duals of $\cn =1$ chiral gauge theories 
in the Coulomb phase, near a strongly coupled superconformal fixed point. The gauge/string dual pair that we consider here 
provides  an example of $\cn = 1$ holography that can be trusted even for a small number 
of colors --~in opposition with \textsc{AdS}$_5$/\textsc{cft}$_4$ when one needs to take 
a 't Hooft large $N$ limit because only the supergravity 
regime is available~-- and the worldsheet instantons effects are fully included 
in the abstract algebraic description of the worldsheet theory. Indeed the string theory vacua, 
being obtained as an asymmetric orbifold of the $\cn = 2$ model, is still free of Ramond-Ramond 
fluxes, and the worldsheet theory still exactly 
solvable.

The type of 
gauge theories that we will obtain are quiver gauge theories, i.e. arrays of 
gauge groups with bifundamental matter. Theories of this sort have been introduced  
some time ago~\cite{Halpern:1975yj}, an appeared more recently in the context 
of D-branes on orbifold singularities~\cite{Douglas:1996sw}. The particular models 
considered in this paper were first studied in~\cite{Csaki:1997zg}, and then obtained 
in string theory using an NS5-brane/D-brane/orbifold 
setup~\cite{Lykken:1997gy}. They can be also 
found using a T-dual "brane box" model~\cite{Hanany:1997tb}, as well as 
some D-branes on singularities constructions~\cite{Hanany:1998it}. 
These field theories have been found also to be useful to "deconstruct" 
higher dimensional gauge theories~\cite{Arkani-Hamed:2001nc}.

We will study the gauge 
theory in the semi-classical regime --~using a boundary worldsheet
\textsc{cft} construction involving D-branes suspended between NS5-branes at 
orbifold singularities~-- and then identify the $\cn = 1$ 
four-dimensional non-critical string related by holography to an
Argyres-Douglas superconformal field theory that can be found in the 
moduli space of the Coulomb phase of the quiver. 
We give the explicit mapping between the gauge-theory chiral operators and
massless string states near the superconformal fixed point 
of the quiver theory. The monopoles and dyons that would become massless 
at the superconformal fixed point correspond in the string theory 
to non-supersymmetric D-branes, which are nevertheless stable.

We start in section~\ref{secN2lst} by reviewing the construction of $\cn = 2$ 
non-critical superstrings in four dimensions, giving their partition function and the massless 
spectrum. In section~\ref{N2dual} we discuss the duality with 
$\cn =2$ gauge theories near a superconformal fixed point and study the localized D-branes  
corresponding to the light dyons. Then in section~\ref{n1lst} we explain how to 
construct four-dimensional vacua of type \textsc{ii} non-critical strings 
with $\cn = 1$ supersymmetry. The dual gauge theory is discussed 
in section~\ref{n1duals}, as well as the localized D-branes. 
Finally in the conclusion we discuss some generalizations 
that will be the object of a forthcoming paper. In the appendix are gathered some facts 
about characters of $\cn = 2$ worldsheet superconformal field theories. 

%%%%%%%%%%%%%%%%%%%%%%%%%%%%%%%%%%%%%%%%%%%%%%%%%%%%%%%%%%%%%%%%%%%%%%%%%%%%%%%%%%%%%%%%%%%%%
%%%%%%%%%%%%%%%GENERAL ASPECTS OF 4D LST%%%%%%%%%%%%%%%%%%%%%%%%%%%%%%%%%%%%%%%%%%%%%%%%%%%%%
%%%%%%%%%%%%%%%%%%%%%%%%%%%%%%%%%%%%%%%%%%%%%%%%%%%%%%%%%%%%%%%%%%%%%%%%%%%%%%%%%%%%%%%%%%%%%
\boldmath
\section{Four dimensional non-critical superstrings with $\cn = 2$ supersymmetry}
\unboldmath
\label{secN2lst}
Four dimensional non-critical superstrings are defined~\cite{Giveon:1999zm} as $\cn=2$ linear dilaton 
backgrounds with four-dimensional Poincar\'e invariance. A large class of those can be built out of 
\begin{equation}
\er^{3,1} \ \times \ \er_{Q} \ \times \ U(1) \ \times \left. \frac{SU(2)}{U(1)} \right|_{k_1} 
\ \times \  \left. \frac{SU(2)}{U(1)} \right|_{k_2} \, ,
\label{back4lst}
\end{equation}
where $k_1$ and $k_2$ are arbitrary integers bigger or equal to two, corresponding to the levels of the 
super-cosets $SU(2)/U(1)$, the minimal models of the $\cn = 2$ superconformal algebra. 
The background charge $Q$ of the linear dilaton is fixed by the cancellation of the conformal 
anomaly to be
\begin{equation}
Q = \sqrt{\frac{1}{k_1}+\frac{1}{k_2}}
\end{equation}
and the $U(1)$ factor is requested for space-time supersymmetry, since it will play the role of the 
$U(1)$ R-symmetry in spacetime. 

In this note we will consider the special case $k_2 = 2$ and $k_1 = n$, i.e. with only one minimal model 
(the super-coset SU(2)/U(1) at level two contains just the identity and thus trivializes) 
\begin{equation}
\er^{3,1} \ \times \ \er_{Q} \ \times \ U(1) \ \times \left. \frac{SU(2)}{U(1)} \right|_{n} \, .  
\label{back4lstred}
\end{equation}
The background charge of this linear dilaton background is then given by 
\begin{equation}
Q = \sqrt{\frac{n+2}{n}}\, . 
\end{equation}
Note that the superstring theories~(\ref{back4lstred}) are intrinsically
non-critical and their effective action 
receives large $\alpha'$ corrections. In the special case $n = 2$ 
there are no $SU(2)/U(1)$ factors  and the string theory 
is the well-known worldsheet \textsc{cft} mirror to the conifold~\cite{Mukhi:1993zb,Ghoshal:1995wm}.
We will study the more generic situation~(\ref{back4lst}) in a following publication.

To construct a spacetime-supersymmetric background out 
of these $\cn=2$ \textsc{scft} building blocks we need to perform a generalized \textsc{gso} projection, 
i.e. to keep only states (in the light-cone gauge) 
with odd-integral left and right worldsheet R-charge~\cite{Gepner:1987qi}. 
Before this orbifoldization, the left and right R-charges are given as follows 
\begin{eqnarray}
Q_R & = & Q_\textsc{fer} + \sqrt{1+\frac{2}{n}}\  
p^\textsc{x}_\textsc{l} - \frac{m}{n}  \mod 2  \, , \nonumber \\
\bar Q_R & = & \bar Q_\textsc{fer} + \sqrt{1+\frac{2}{n}}\  p^\textsc{x}_\textsc{r} 
+ \frac{\bar m}{n}  \mod 2 \, ,
\label{defrcharge}
\end{eqnarray}
in terms of the left and right momenta $p^\textsc{x}_\textsc{l,r}$ along the $U(1)$ free field $X$, and 
$m/2, \, \bar m/2$  the left and right $U(1) \subset SU(2)$ eigenvalues 
in the $\cn=2$ coset $SU(2)/U(1)|_{n}$. Note that $\bar m$ enters in the R-charge 
of the right-movers with opposite sign compared to $m$ for the left-movers. 
The charges $Q_\textsc{fer}$ and $\bar Q_\textsc{fer}$ 
are the ordinary contributions form the fermion number in the \textsc{ns} or in the \textsc{r} sector. 
Clearly the compact boson $X$ has to be compactified to a specific radius and twisted to obtain a 
\textsc{gso}-projected theory.

\subsection{The one-loop partition function}
The information about the spectrum can be read out of the one-loop vacuum amplitude. A consistent spectrum 
has to be compatible with the modular invariance of this partition function on the torus. We will 
concentrate on the contribution to the partition function of delta function-normalizable operators, i.e. 
those which propagate along the non-compact linear dilaton direction~$\rho$,  
\begin{equation}
\mathcal{O}_P (z, \bar z ) = e^{-Q(\frac{1}{2}+i  P )\rho (z, \bar z)} \ \mathcal{Y} (z , \bar z)\, ,
\end{equation}
where $\mathcal{Y} (z, \bar z)$ is the vertex operator corresponding to the other degrees of freedom, 
including the superconformal ghosts. The non-normalizable operators are also important and we will discuss 
them below, however they don't enter (by construction) in the one-loop vacuum amplitude. 

The spectrum of the delta-normalizable operators is insensitive to the physics in the strong coupling 
region $\rho \to -\infty$, in particular to the presence of 
a potential that would regularize it (as we will consider later), 
even though the sub-leading term in the density of states 
depend upon this potential, as can be shown using a path-integral approach 
to the partition function~\cite{Maldacena:2000kv,Hanany:2002ev,Eguchi:2004yi,Israel:2004ir}. 
We are interested in the following in the type \textsc{iia} superstring 
theory, for which the four-dimensional non-critical string 
will be dual to $\cn =2$ gauge theory in a way that we will discuss below. 
The modular-invariant partition function for the four-dimensional non-critical strings 
studied in this paper is given by (see also~\cite{Eguchi:2004yi}) 
\begin{eqnarray}
Z_\textsc{cont} (\tau, \bar \tau ) 
&=&  \frac{1}{4\pi^2 \alpha' \tau_2} \frac{1}{\eta^2 (\tau ) \bar \eta^2 (\bar \tau )}\ \frac{1}{4}\!\!\!\!  \sum_{a,b,\bar a , \bar b \in \zi_2} 
\sum_{\{ \upsilon_\ell \} , \{ \bar \upsilon_\ell \} \in (\zi_2)^4}(-)^{a+\bar a + 
b (1 + \sum_i \upsilon_i) + 
\bar b (1 + \sum_i \bar \upsilon_i)+\bar a \bar b} \ \times 
\nonumber \\ & \times &  
\frac{\Theta_{a+2\upsilon_1,2} (\tau ) \ \bar 
\Theta_{\bar a+2\bar \upsilon_1,2} (\bar \tau )}{\eta (\tau ) \bar 
\eta (\bar \tau ) }
  \sum_{2j =0}^{n-2}  \sum_{m , \bar m \in \zi_{2n}} 
 C^{j\, (a+2\upsilon_3)}_{m} (\tau )\   
\bar C^{j\, (\bar a+2\bar \upsilon_3)}_{\bar   m} (\bar \tau )\  \times   \nonumber \\
& \times &  
\int_0^\infty \di P
\sum_{ r \in \zi_{n + 2} }  
Ch_c^{(a+2\upsilon_2)} (P,2 m + n (a+2\upsilon_4)+ 4 n r; \tau  ) \ \times
\nonumber \\
& \times &  
\bar{Ch}_c^{(\bar a+2\bar \upsilon_2)} (P,-2 \bar m + n (\bar a+2\bar \upsilon_4)+
4 n \bar r; \bar \tau )\    
\delta_{m -a-2\upsilon_4-4 r ,\bar m +\bar a+2\bar \upsilon_4 +4 \bar r \!\!\!  \mod
  2(n + 2)}\nonumber \\
\label{partcontred}
\end{eqnarray}
One can check that this one-loop amplitude is modular-invariant and has the required 
properties (spin-statistics, worldsheet local $\cn= 1$ superconformal symmetry and 
locality w.r.t. the spacetime supercharges). The partition function is 
expressed in terms of $C^{j\, (s)}_{m}$, the  characters of the $\cn = 2$ minimal model 
$SU(2)/U(1)|_{n}$. The contribution of the linear dilaton direction and the compact U(1) free 
field $J = i \p X$, as well as their fermionic superpartners, 
have been recast together in terms of $Ch_c^{(s)} (P,m )$, the extended 
characters of $\cn = 2$ superconformal algebra with $c= 3 + 3 Q^2$. The main properties of these 
characters are gathered in the appendix. The left and right worldsheet R-charges 
of this model, defined by eq.~(\ref{defrcharge}), are integral for all the states propagating in this 
one-loop amplitude, and the final projection giving a set of mutually local
vertex operators is ensured by the $\zi_2$ projection acting on the fermion number, 
similar to superstrings in flat ten-dimensional spacetime. 
Thus this is a spacetime supersymmetric model, with one supersymmetry from 
the left-movers and one from the right-movers. To write explicitly the spacetime supercharges we can bosonize the 
worldsheet fermions of  the $\mathbb{R}_Q \times U(1)$ theory as
\begin{equation}
\xi^\pm (z) = \frac{
\xi^\rho (z) \pm i\xi^\textsc{x} (z)}{\sqrt{2}} = e^{\pm i H_2 (z)}  \quad , \qquad 
\tilde \xi^\pm (\bar z) = \frac{
\tilde \xi^\rho (\bar z)\pm i \tilde \xi^\textsc{x} (\bar z)}{\sqrt{2}} = e^{\pm i \tilde H_2  (\bar z)} 
\, .
\end{equation}
Then the spacetime supercharges of the gravitational theory come from the
gravitino vertex operators. 
However when one considers the non-gravitational dual of this string theory it is natural to 
consider instead the four-dimensional supercharges as given by (in the $(-\nicefrac{1}{2},0)$ picture)
\begin{eqnarray}
\mathcal{Q}^\textsc{l}_{\alpha} (z) & = & \oint \di z \  e^{-\frac{\phi (z)}{2}}  \
\mathcal{S}^\textsc{l}_\alpha (z)\   
e^{+ i \frac{Q}{2} \, X_L (z)} \ e^{+\frac{i}{2}  H_2 (z)}\  
V_{0,\frac{1}{2},0}^{(1,0)} (z,\bar z)\, ,  \nonumber \\
\bar{\mathcal{Q}}^\textsc{l}_{\dot \alpha}( z) & =&  \oint \di z \ e^{-\frac{\phi (z)}{2}}  \
\bar{\mathcal{S}}^\textsc{l}_{\dot \alpha} (z) \  
e^{ -i \frac{Q}{2} \, X_L (z)} \ e^{-\frac{i}{2}  H_2 (z)}\  
V_{0,-\frac{1}{2},0}^{(3,0)} (z,\bar z) \, , \nonumber
\label{stsupercharges}
\end{eqnarray}
in terms of the $SL(2,\mathbb{C})$ spin fields 
$\mathcal{S}^\textsc{l}_{\alpha}$ and $\bar{\mathcal{S}}^\textsc{l}_{\dot \alpha}$ 
from the left-moving worldsheet \textsc{scft}. We denote by $V^{(s,\bar s)}_{j, m/2, \bar m/ 2}$ 
a primary of the coset $[SU(2)_{n-2} \times SO(2)_1]/U(1)$ with quantum numbers $(j,m, \bar m, s, \bar s)$, 
see the appendix for conventions. 
These supercharges are non-normalizable operators in the bulk. 
The spacetime supercharges from the right-moving sector are constructed in the same manner: 
\begin{eqnarray}
\mathcal{Q}^\textsc{r}_{\alpha}( \bar z) & = & \oint \di \bar z \  e^{-\frac{\tilde \phi (\bar z)}{2}}  \
\mathcal{S}^\textsc{r}_\alpha (\bar z)\   
e^{ - i\frac{Q}{2} \, X_R (\bar z)} \ e^{-\frac{i}{2}  \tilde{H}_2 (\bar z)}\  
V_{0,0,+\frac{1}{2}}^{(0,3)} (z,\bar z) \, , \nonumber \\
\bar{\mathcal{Q}}^\textsc{r}_{\dot \alpha} (\bar z)
& = & \oint \di \bar z \  e^{-\frac{\tilde \phi (\bar z)}{2}}  \
\bar{\mathcal{S}}^\textsc{r}_{\dot \alpha} (\bar z)\   
e^{+i \frac{Q}{2} \, X_R (\bar z)} \ e^{+\frac{i}{2}  \tilde{H}_2 (\bar z)}\  
V_{0,0,-\frac{1}{2}}^{(0,1)} (z,\bar z)\, ,  \nonumber \\
\label{stsuperchargesr}
\end{eqnarray}
with now the spin fields from the right-movers. Overall it leads to $\cn=2$
supersymmetry in four dimensions.

%%%%%%%%%%%%%%%%%%%%%%%%%%%%%%%%%%%%%%%%%%%%%%%%%%%%%%%%%%%%%%%%%%%%%%%%%%%%%%%%%%%%%%%%%%%%%%%%%%%%%%%%
%%%%%%%%%%%%%%%%%%MASSLESS SPECTRUM AND RESOLUTION OF THE SINGULARITY%%%%%%%%%%%%%%%%%%%%%%%%%%%%%%%%%%%
%%%%%%%%%%%%%%%%%%%%%%%%%%%%%%%%%%%%%%%%%%%%%%%%%%%%%%%%%%%%%%%%%%%%%%%%%%%%%%%%%%%%%%%%%%%%%%%%%%%%%%%%

\subsection{Massless spectrum}
\label{speclst} 
\label{nschir}
In order to obtain precise statements about the holographic duals of these models, 
we will study the spectrum of string operators which are massless in space-time. 
The allowed quantum numbers for the 
minimal models $SU(2)/U(1)|_{n}$  and for the $U(1)$ compact boson 
can be read from the partition function given 
by eq.~(\ref{partcontred}).

We start with the massless states of this type \textsc{iia} non-critical 
superstring in the \textsc{ns}-\textsc{ns} sector. First we find 
the universal hypermultiplet and the graviton multiplet from the 
delta-function normalizable operators of the linear dilaton. 
Then, we are looking for massless states with zero worldsheet R-charge from the four-dimensional 
spacetime part, that are non-normalizable along the linear dilaton 
direction. These operators have support in the weak coupling region, and are
believed to be dual to off-shell, 
gauge-invariant operators in the four-dimensional gauge theory. 
We need to glue the massless states from the left- and right-moving sectors  in a way compatible 
with the one-loop amplitude~(\ref{partcontred}). 
We can construct worldsheet chiral primaries of dimension $h = Q/2=1/2$ by
starting with a chiral primary of 
$SU(2)/U(1)|_{n}$ of dimension $h = \nicefrac{1}{2} - \nicefrac{j+1}{n}$,  coming from a state in an 
$SU(2)$ representation of spin $j$ 
with $m = 2(j + 1)$ and $s_3=2$, i.e. fermion number one. 
For the right moving sector we can choose an anti-chiral 
primary with $(\bar m = 2(j + 1),\bar s_3 = 2)$. Such 
a state is compatible with the partition function and corresponds to the  
vertex operator (in the $(-1,-1)$ picture): 
\begin{equation}
\mathcal{V}^{(c,a)}_j = 
e^{-\varphi-\tilde \varphi}\ e^{i p_\mu X^\mu} \ e^{-Q \tilde \jmath  \rho} \ 
e^{i \frac{2 (j + 1)}{\sqrt{n (n + 2)}} (X_L - X_R)} \
V_{j, \, j+1, \, j +1}^{(2,2)}\, .
\label{vca}
\end{equation}
One can first define this massless operator on-shell, from the four-dimensional 
point of view, in a way giving a worldsheet chiral $(c,a)$ primary, 
if we choose for the linear dilaton momentum 
\begin{equation} 
\tilde \jmath  =  \frac{2 (j+1)}{n+ 2}\, .
\label{secbranchv}
\end{equation}
Then it will be dual to an operator of the dual theory if it obeys the 
Seiberg bound~\cite{Seiberg:1990eb}, i.e. if it is non-normalizable:
\begin{equation}
\tilde \jmath < \frac{1}{2} \implies 4 (j + 1)  < n+2 \, . 
\end{equation}
By taking values of $\tilde \jmath$ different from~(\ref{secbranchv}) this operator 
can be defined off-shell (in the four-dimensional sense). 
There is another dressing of these operators, giving a physical state 
but which is not an worldsheet chiral primary, obtained by replacing  $\tilde
\jmath  \to 1 - \tilde  \jmath $ 
and with a Seiberg bound now given by 
\begin{equation}
\tilde \jmath  = \frac{n  -  2j}{n + 2} \ \implies \ 2 j > \frac{n}{2} + 1 \, . 
\label{fibranchv}
\end{equation} 
These massless operators are chiral in spacetime. From the worldsheet 
expression of the space supercharges~(\ref{stsupercharges}) we find that 
they are all bottom components of  $\cn =2$ vector multiplets in spacetime. 
We can of course construct similarly $(a,c)$ primary states. However 
there are no massless states in the spectrum that 
are worldsheet $(c,c)$ or $(a,a)$ primaries, as was already observed in~\cite{Eguchi:2004ik}, 
so no hypermultiplets in the dual gauge theory that are mapped to closed string states. 
The R-charge in spacetime of the $(c,a)$ and $(a,c)$ operators that we found can be defined as 
\begin{equation}
R =   \frac{2i}{Q} \oint (\p X  - \pb X)  =  \frac{8 (j + 1) }{n + 2}\, .
\label{rch}
\end{equation}
The R-charge is thus identified with the winding number $w$ around the $U(1)$, as 
$w =Q^2 R /4$. In the double scaling 
limit, this will correspond to winding around the cigar, which is not conserved. It 
is consistent with the breaking of the $U(1)_R$ R-symmetry in the gauge theory
dual which is not superconformal in the double scaling limit.\footnote{
This convention is consistent with the fact that, 
with the selection rule that appears in the partition
function~(\ref{partcontred}), only the $(c,a)$ and the $(a,c)$ states are part of the spectrum. 
One could instead construct a consistent model if we kept only $(c,c)$ and $(a,a)$; 
these two choices correspond to  
T-dual models. Our choice for the R-charge~(\ref{rch}) as a winding number is consistent 
with keeping only $(c,a)$ and $(a,c)$ states.} Also the Cartan eigenvalue of the 
$SU(2)_R$ symmetry is given by the momentum through, for a generic operator 
\begin{equation}
m_\textsc{r} = \frac{i}{Q} \oint (\p X + \pb X) = \frac{a+\bar a}{2} +
\upsilon_4 + \bar \upsilon_4 \mod 2   \, .
\end{equation} 
This charge spectrum is similar to the Cartan of a diagonal $SO(3)$ 
from the $SO(3)_1 \times SO(3)_1$ algebra of 3 left-moving and 3 right-moving fermions; 
however the other $SO(3)$ generators are not affine symmetries of the worldsheet theory. 
The \textsc{ns-ns} chiral operators that we are considering are singlets of this $SU(2)_R$, as
expected  since they are bottom component of  vector multiplets. 

It is expected on general grounds that these string operators, that are massless, chiral in spacetime 
and not normalizable (i.e. they satisfy the Seiberg bound), are dual to chiral off-shell operators in the 
dual non-gravitational theory. Giving a (constant) vacuum expectation value
(\textsc{vev}) to these operators should be achieved by taking the {\it normalizable branch} 
of the same operators, i.e. the branch violating the 
Seiberg bound, at zero spacetime momentum. Then, since no spacetime supersymmetry should be broken by
these \textsc{vev}s, the corresponding worldsheet operator has to be
$\cn = 2$ worldsheet chiral. These considerations show that the dual chiral operators 
in spacetime are given (on-shell) by the branch~(\ref{fibranchv}), and their \textsc{vev}s are given by 
considering the operators in the branch~(\ref{secbranchv}) violating the Seiberg bound, i.e. for 
the $SU(2)$ spins $j> \nicefrac{n-2}{4}$.

For completeness we study the operators the \textsc{r}-\textsc{r} sector. We have to 
look simply for the Ramond ground states of each of the $\mathcal{N}=2$ \textsc{scft} factors 
of the string theory. Let's take to begin an $SU(2)/U(1)_{n}$ coset. 
The Ramond ground states are given by $(m = 2j+1, \upsilon_3=0)$ 
or  $(m =- 2j-1, \upsilon_3=1)$ which 
are respectively the one-half spectral flow of a chiral primary and of an anti-chiral one. 
For the  $[\mathbb{R}_Q \times U(1) ]$ factor one can have also a Ramond ground state for 
$\upsilon_2=0$ if $p_\textsc{x}  >0 $, or and $\upsilon_2=1$ if $p_\textsc{x} <0$, which would correspond 
respectively to the one-half spectral flow of a chiral primary and of an anti-chiral one, provided 
we choose $\tilde \jmath$ accordingly. However not all the combinations appear 
in the spectrum of string theory. 
As for the supercharges are given by~(\ref{stsupercharges}), only the operators 
with $\upsilon_2=\upsilon_3=\upsilon_4$ 
(and similarly $\bar \upsilon_2=\bar \upsilon_3=\bar \upsilon_4$) appear. Then we obtain massless states in the string theory 
given by (in the $(-\nicefrac{1}{2},-\nicefrac{1}{2})$ picture)
\begin{eqnarray}
\mathcal{U}_{j} = C_{\mu \nu}
 e^{-\frac{\phi  + \tilde \phi}{2}}  e^{-Q\tilde \jmath \rho}
\left\{ 
\mathcal{S}^\textsc{l}_{\alpha} \sigma^{\mu \nu \ \alpha}_{\quad \ \beta}\,  \mathcal{S}^{\textsc{r} \ \beta} 
e^{ i Q\left[\frac{2j}{n+2}+\frac{1}{2}\right]( X_L (z)-X_R (\bar z) )}
\ e^{\frac{i}{2}  (H_2 (z)- \tilde H_2 (\bar z))}\  
V_{j,\, j+\frac{1}{2},\, j+\frac{1}{2}}^{(1,3)} \right. \nonumber \\
\left. +  \ \  
\bar{\mathcal{S}}^\textsc{l}_{\dot \alpha} \bar{\sigma}^{\mu \nu \ \dot \alpha}_{\quad \ \dot \beta}\,  
\bar{\mathcal{S}}^{\textsc{r} \ \dot \beta} 
e^{ -i Q\left[\frac{2j}{n+2}+\frac{1}{2}\right]( X_L (z)-X_R (\bar z) )}
\ e^{-\frac{i}{2}  (H_2 (z)- \tilde H_2 (\bar z))}\  
V_{j,\,-j-\frac{1}{2},\,-j-\frac{1}{2}}^{(3,1)}
\right\} \, , \nonumber \\
\end{eqnarray}
where $C_{\mu \nu}$ is a polarization for the field strength 
of the vector field. These operators 
are indeed \textsc{r-r} ground states if we choose for the Liouville momentum 
$\tilde \jmath =  \frac{2j}{n + 2}$. The self- and anti-self-dual parts have  $U(1)_R$ charge 
\begin{equation}
R =\pm \left[ 2+ \frac{8j}{n+2} \right] \, , 
\end{equation}
and are singlets of $SU(2)_R$, as expected.

\subsection{Resolution of the singularity}
To obtain a manageable perturbative description of this non-critical string, one 
needs to regularize the strong coupling of the linear dilaton. 
From the worldsheet point of view it amounts to add an appropriate potential to the worldsheet action, such 
that the D-branes localized in the strong coupling region become all massive. 

We expect that the worldsheet chiral operators that we discussed that obey the Seiberg bound will be dual to 
chiral operators in the spacetime holographic theory. They have support in the weak coupling region 
$\rho \to \infty$. We have seen that in our four-dimensional \textsc{lst} the 
\textsc{ns}-\textsc{ns} physical operators that satisfy the Seiberg bound 
are given by~(\ref{vca}). Giving \textsc{vev}s to these operators can be achieved, as we 
discussed above, if one adds to the worldsheet action the conjugate of such an operator 
under $\tilde \jmath \to 1-\tilde \jmath \, $, 
which has support in the strong coupling region $\rho \to - \infty$ and violates the Seiberg bound. 
For such an operator not to break spacetime supersymmetry it has to be 
a worldsheet chiral operator of the $\cn =2$ superconformal algebra. 
The first example from~(\ref{vca}) is the "$\mathcal{N}=2$ Liouville operator"  
\begin{equation}
\mathcal V^{(c,a)}_{\frac{n}{2}-1} = 
e^{-\varphi-\tilde \varphi}\ e^{ip_\mu X^\mu }\ e^{-Q\tilde \jmath \rho} e^{i \sqrt{\frac{n}{ n + 2}} (X_L - X_R)} 
V_{\frac{n}{2}-1 , \frac{n}{2} , \frac{n}{2}}^{(2,2)} \, . 
\end{equation}
Giving a vacuum expectation value to the dual of this operator in the spacetime theory  
is then achieved by adding this operator in the first branch~(\ref{secbranchv}), 
for which it is a worldsheet $(c,a)$ state of the $\cn = 2$ algebra, in the $(0,0)$ picture (and its conjugate):
\begin{equation}
\delta \mathcal{L}= \mu_\textsc{Liouville}  G_{-\nicefrac{1}{2}}^- \tilde G_{-\nicefrac{1}{2}}^+
\ e^{ -\sqrt{\frac{n}{n + 2}} \left[ \rho - i(X_L - X_R)\right]} \ + \ \text{c.c.} \, , 
\end{equation}
using the field identification in $\cn = 2$ minimal models 
$(j,m,s) \sim (\frac{n}{2}-j-1,m+n,s+2)$.

After turning on this exactly marginal worldsheet deformation, the strong coupling region 
is regularized in a way that preserve $\cn = 2$ superconformal symmetry. Indeed, if 
we build an $\cn = 2$ worldsheet twisted chiral superfield whose bottom
component is $\phi = \rho - i (X_L - X_R)$, we can write this potential as a twisted F-term 
\begin{equation}
\delta \mathcal{L} =  \mu_\textsc{Liouville}\ \int \di \bar \theta^- \di  \theta^+ \ 
e^{ - \frac{\Phi}{Q}} \ + \ \text{c.c.} \, .
\label{n2pot}
\end{equation}

There is another type of exactly marginal worldsheet perturbation, preserving $\cn =2$ superconformal 
symmetry on the worldsheet, that can be added to the Lagrangian. We can consider the 
following dressing of the identity 
in the $\cn = 2$ minimal models:
\begin{equation}
\delta \mathcal{L}= \mu_\textsc{cigar} \  G_{-\nicefrac{1}{2}}^+ \tilde G_{-\nicefrac{1}{2}}^+ 
\ \xi^- \tilde{\xi}^- \ e^{ -\sqrt{\frac{n+2}{ n}}  \rho } \, . 
\label{cigpert}
\end{equation}
The $(0,0)$ picture vertex operator added to the Lagrangian 
is then a supersymmetric descendant of a worldsheet $(a,a)$ operator.
However in opposition with the previous type of operators, it is physical only 
for the specific value of the Liouville momentum $\tilde \jmath  =1$, 
which violates the Seiberg bound;  thus 
this normalizable operator cannot be viewed as holographically dual of a vacuum expectation value 
for some gauge theory observable. 
It we consider the theory already perturbed by the $\cn = 2$ potential, see eq.~(\ref{n2pot}), 
we can nevertheless consider a normalizable operator with those quantum numbers. 
It is so possible to turn on a marginal perturbation with this asymptotic expansion. 
As for the $\cn = 2$ Liouville potential, it can be also written 
in terms of the $\cn =2$ twisted chiral superfield as a D-term 
\begin{equation}
\delta \mathcal{L}= -\mu_\textsc{cigar} \ \int \di^4 \theta \ e^{ -\frac{Q}{2}  
(\Phi+\Phi^*)}
\end{equation}
Up to total derivatives, this operator is the first term in the asymptotic expansion of the 
supersymmetric sigma model with a cigar geometry
\begin{equation}
\di s^2 = \di \rho^2 + \tanh^2 (\nicefrac{\rho }{\sqrt{\alpha' k}}) \ \di X^2 \, , 
\end{equation}
with $k = \frac{2n}{n+2}$. 
This cigar geometry is the target space metric of the gauged super-\textsc{wzw} model $\slc$ at 
level $k$~\cite{Elitzur:1991cb,Mandal:1991tz,Witten:1991yr}. This coset model has also a 
non-trivial dilaton profile, which is bounded from above and asymptotically linear 
\begin{equation}
\Phi = \Phi_0 + \frac{1}{2}\log k - \log \cosh (\nicefrac{\rho}{\sqrt{\alpha' k}}) \, , 
\end{equation}
such that the coefficient of the perturbation~(\ref{cigpert}) of the linear
dilaton background can be related 
to the zero mode of the dilaton in the cigar theory as
$\mu_\textsc{cigar} = 4 e^{-2\phi_0}/k$. 
Consequently the coefficient of the cigar perturbation is related to the string coupling at the tip of the
cigar 
\begin{equation}
g_\textsc{eff}^2 = k e^{2 \Phi_0}=\frac{4}{\mu_\textsc{cigar}} \, .
\label{cigcoupl}
\end{equation}
However we should be aware that, because of the \textsc{gso} projection, the 
cigar geometry is not meaningful in the four-dimensional non-critical string
that we consider; in particular, the 
background for the target space contains non-zero flux for the \textsc{ns-ns} 
two-form since it represents a wrapped NS5-brane.  

To summarize we can regularize
the strong coupling regime of the linear dilaton by adding the $\mathcal{N}=2$
Liouville potential and/or the cigar
perturbation. It is known~\cite{Giveon:2001up} 
that a the consistent $\slc$ \textsc{scft} requires that both perturbations are present at the same 
time, with related coefficients (we will come back later to this important issue). Then we consider 
our four-dimensional non-critical strings with the replacement
$\mathbb{R}_{Q} \times S^1_\textsc{x} \longrightarrow \slc |_{\frac{2n}{n+2}}$, 
giving the double scaling limit of four-dimensional little string
theory~\cite{Giveon:1999px}. In the $\slc$ \textsc{cft}, the winding around the compact
direction is not conserved; this intuitive fact is related algebraically to 
the fact that free-field computations of the correlators involve the 
insertion of the $\cn =2$ Liouville interaction~(\ref{n2pot}) 
as screening charges, thus violating winding conservation by integer 
amounts. However in our string construction, because of the 
\textsc{gso} projection, the winding is fractional, leaving a conserved 
$\zi_{n}$ charge. This will be matched precisely with the $U(1)_R$ anomaly 
of the gauge theory. Note finally that they are various more complicated 
regularizations of the strong coupling regime (by turning on other
deformations), however we don't have an explicit solution of the 
resulting worldsheet conformal field theory because they have a non-trivial $SU(2)/U(1)$ dependence. 

\subsection{Discrete spectrum in the double scaling limit}
As we already mentioned, this $\slc$-based background contains not only delta-norma\-lizable 
states, given by the partition function~(\ref{partcontred}), and non-normalizable states 
discussed above, but also a discrete spectrum of normalizable 
states. They can be thought as bound states living near the tip of the cigar.  
The contribution of these discrete states is requested 
for modular-invariance of the partition function and reads 
(see~\cite{Eguchi:2004yi,Israel:2004ir,Eguchi:2004ik} for more details):
\begin{eqnarray}
Z_\textsc{disc} = 
\frac{1}{4\pi^2 \alpha' \tau_2} \frac{1}{\eta^2 \bar \eta^2 }\ \frac{1}{4}\!\!\!
\sum_{\{ \upsilon_\ell \} , \{ \bar \upsilon_\ell \} \in (\zi_2)^4}
 \sum_{a,b,\bar a , \bar b} (-)^{a+\bar a + 
b (1 + \sum_i \upsilon_i) + 
\bar b (1 + \sum_i \bar \upsilon_i) +\bar a \bar b} 
\frac{\Theta_{a+2\upsilon_1,2} \Theta_{\bar a+2\bar \upsilon_1,2}}{\eta \bar \eta}  
\nonumber \\ 
\sum_{2j =0}^{n-2}
\sum_{m , \bar m \in \zi_{2n}}
\sum_{ r \in \zi_{n + 2} } \ \int_{\frac{1}{2}}^{\frac{n}{n+2} +\frac{1}{2}} \di \tilde \jmath \quad 
\delta \left( \tilde \jmath -\frac{m-\bar m +2n(r+\bar r )-a-\bar a +n(\upsilon_4+\bar \upsilon_4)}{2(n+2)} + \frac{\zi}{2} \right) \nonumber \\
Ch_d^{(a+2\upsilon_2)} (\tilde \jmath,2 m + n (a+2\upsilon_4)+ 4 n r )
\bar{Ch}_c^{(\bar a+2\bar \upsilon_2)} (\tilde \jmath,-2 \bar m + n (a+2\bar \upsilon_4)+ 4 n \bar r )
\ \times \nonumber \\ \times \ 
 \delta_{m -a-2\upsilon_4-4 r ,\bar m +\bar a+2\bar \upsilon_4 +4 \bar r   \mod 2(n + 2)}\ \ 
 C^{j\, (a+2\upsilon_3)}_{m} \  \bar C^{j\, (\bar a+2\bar \upsilon_3)}_{\bar m}\nonumber \\
\label{partdisc}
\end{eqnarray}
written in terms of the discrete extended characters of the supersymmetric $\slc$ coset 
$Ch_d^{(s)} (j,m)$ defined in the appendix. Thanks to the $\delta$-function that implements 
the constraint of the gauging the spectrum of representations of $\slr$ spin $\tilde \jmath$ is discrete. 
In the massless sector it contains the normalizable states that we expect from our previous analysis, that 
correspond to \textsc{vev}s for the observables dual to the worldsheet operators~(\ref{vca}).

%%%%%%%%%%%%%%%%%%%%%%%%%%%%%%%%%%%%%%%%%%%%%%%%%%%%%%%%%%%%%%%%%%%%%%%%%%%%%%%%%%%%%%%%%%
%%%%%%%%%%%%%%%%%%%%%%%%%%%%%%%%%%%%%%%%%%%%%%%%%%%%%%%%%%%%%%%%%%%%%%%%%%%%%%%%%%%%%%%%%%%
%%%%%%%%%%%%%%%%%%%%PURE SU(N) AND DUALITY%%%%%%%%%%%%%%%%%%%%%%%%%%%%%%%%%%%%%%%%%%%%%%%%%
%%%%%%%%%%%%%%%%%%%%%%%%%%%%%%%%%%%%%%%%%%%%%%%%%%%%%%%%%%%%%%%%%%%%%%%%%%%%%%%%%%%%%%%%%%%

\boldmath
\section{Holographic duals of $\cn = 2$ non-critical superstrings}
\unboldmath
\label{N2dual}
In this section we will discuss the holographic interpretation of these 
3+1-dimensional non-critical superstrings.   
We will focus on the examples corresponding to the $A_n$ modular invariant (i.e. 
diagonal w.r.t. $j$) for the $\cn= 2$ minimal model 
$SU(2)/U(1)$ that enters in the string theory description detailed in the previous section. 
One can of course also consider the D and E series in a similar fashion. 

\subsection{Argyres-Douglas superconformal field theories and wrapped fivebranes}
The low-energy effective action of an $\cn = 2$ $SU(n)$ gauge theory at a 
generic point in the moduli space of the Coulomb branch contains $(n-1)$ 
Abelian vector multiplets of $\cn = 2$. 
The moduli space is parameterized by the symmetric polynomials $s_\ell$ 
in the eigenvalues of $\phi$, the scalar component of the gauge multiplet, obtained by the generating 
function
\begin{equation}
P_n(x) : = 
\langle \det ( x -  \phi  )\rangle = \sum_{\ell = 0}^n  s_{\ell} \, x^\ell
\label{polchar}
\end{equation}
with $s_n = 1$ and $s_{n-1} = 0$ due to the tracelessness condition. 
In terms of those the couplings of the exact low energy effective action can
be obtained from the periods of the 
the Seiberg-Witten curve (\textsc{sw})~\cite{Seiberg:1994rs} of the gauge
theory, given by~\cite{Klemm:1994qs,Argyres:1994xh}
\begin{equation}
\Sigma \, : \quad y^2 = \frac{1}{4} P_n (x)^2 -\Lambda^{2n}
\label{swcurve}
\end{equation}
where $\Lambda$ is the dynamically generated scale. 
At some points the moduli space develops singularities, corresponding to new states 
becoming massless, magnetic monopoles and/or dyons. They enter in the effective action as 
hypermultiplets charged under some of the U(1) factors. 
In particular one can get a superconformal field theory if we choose the vacua given by 
\begin{equation}
P_n (x) =  x^n +  v_c \quad \text{with} \qquad v_c  = (-)^n \langle \det \phi \rangle  =\pm 2\Lambda^n    
\end{equation}
called Argyres-Douglas (\textsc{ad}) fixed points~\cite{Argyres:1995jj}. The dyons 
that become massless are mutually non-local and we obtain a strongly coupled $\mathcal{N}=2$  
superconformal theory.  It is the most singular fixed point in all the moduli space of the gauge theory, 
and $n(n-1)/2$ dyons become massless~\cite{Shapere:1999xr}. These 
dyons are charged under only $\lfloor \frac{n-1}{2} \rfloor$ Abelian 
gauge multiplets, thus the additional ones decouple.

The superstring dual of such $\cn = 2$ gauge theories is obtained using a configuration of 
D4-branes suspended between NS5-branes in type IIA superstrings~\cite{Witten:1997sc}, 
that we will discuss in more detail in section~\ref{branesetup} in order to construct 
the $\cn = 1$ models. An $SU(n)$ gauge theory corresponds to a pair of NS5-branes, branes, 
located say at $x^6 = 0$ and $x^6 = L$, with $n$ D4-branes suspended 
between them.\footnote{This configuration leads naively to a $U(n)$ gauge 
theory but the diagonal $U(1)$ --~corresponding to the center 
of mass motion of the D4-branes in the (4,5) plane~-- 
is frozen.} The positions  of the D4-branes in the complex plane 
$x = x^4 + ix^5$ parameterize the Coulomb branch of the theory. 
Then the one-loop effects in the gauge theory correspond to the bending 
of the NS5-branes by the endpoints of the D4-branes (see~\cite{Israel:2005fn} 
for a worldsheet derivation). To include the instantons 
effects, one can lift the brane setup to M-theory, for which the system is described 
by one M5-brane wrapping the Riemann surface of genus $n-1$
\begin{equation}
\Sigma \, : \ t + t^{-1}+ \Lambda^{-2n}\, P_n \left( \frac{2\pi}{\alpha'}\, x \right)  = 0\, ,
\end{equation}
where $t = \exp (- s/R_{10} )$, with 
$s= x^6 + ix^{10}$, and the radius $R_{10}=\sqrt{\alpha '} g_s$ of the eleventh 
dimension gives (in string units) the type IIA string coupling constant.  This curve is exactly  
the Seiberg-Witten curve of the theory, after the change of variables 
$t = \Lambda^{-n}[y - P_n (\frac{2\pi}{\alpha'} x)/2]$.  
When we approach  one of the \textsc{ad} points of the gauge theory the Riemann surface is degenerate; 
thus the eleven-dimensional supergravity picture is not valid. 

Instead we can as in~\cite{Klemm:1996bj} consider a type \textsc{iia} NS5-brane wrapping 
the same curve $\Sigma$; this can be achieved e.g. by compactifying the transverse 
space of the M5-brane along another direction $x^7$, and then performing the 
Kaluza-Klein reduction along this circle rather than along $x^{10}$ (the radius 
of the transverse circle to the NS5-brane becomes irrelevant in the decoupling 
limit that we will consider). The string theory is strongly coupled at the 
Argyres-Douglas point thus the degrees of freedom at the singularity 
can be decoupled from gravity in the limit where the asymptotic string coupling 
constant $g_s$ is sent to zero. From the worldsheet point of view this singular 
wrapped fivebrane solution corresponds to a four-dimensional non-critical 
superstring of the sort~(\ref{back4lstred}), that we studied in detail in the previous section, 
with an infinite throat along the linear dilaton direction. The dyons 
that become massless at the \textsc{ad} point corresponds to 
various D-branes "localized" at infinite string coupling in the 
linear dilaton direction.\footnote{In this strong coupling region one 
would need to come back to the eleven-dimensional picture, 
where we expect to find an AdS$_5$ solution (much like the 5+1-dimensional type \textsc{iia} 
\textsc{lst} that is well described in the deep infrared by $AdS_7 \times S^4$~\cite{Aharony:1998ub}). In 
the present case it is more difficult to make such a picture precise because the 
string background is strongly curved --~and even non-critical. However for the more generic models 
defined in~(\ref{back4lst}) there's a sensible ten-dimensional gravitational interpretation, as we 
will show in a forthcoming publication, and we expect to find an $AdS_5 \times X^6$ solution 
of eleven-dimensional supergravity.} As for the decoupling limit of flat NS5-branes this 
linear dilaton background is holographically dual to the decoupled theory living on 
the singularity; the worldvolume theory on the fivebranes is a non-local little string 
theory~\cite{Seiberg:1997zk} which flows in the infrared for the reasons explained above 
to an Argyres-Douglas superconformal field theory. 

To resolve the singularity corresponding to the Argyres-Douglas fixed point one can turn on \textsc{vev}s 
for the gauge-invariant operators parameterizing the Coulomb branch away from their 
value at the fixed point, in a way giving mass to all the dyons. 
Then we can define a {\it double scaling limit}~\cite{Giveon:1999px} of the dual string theory, 
where the  theory is decoupled from gravity by taking the limit 
$g_s \to 0$, while keeping the masses of the D-branes holographically dual to the 
"light" dyons fixed. This limit keeps only the universal behavior near those 
singularities, i.e. various gauge theories belong to the same universality class; for 
example, the most singular \textsc{ad} fixed point for $SU(n)$ discussed above can be found in 
the moduli space of all $SU(N\geqslant n)$. The most symmetric of those deformations of 
the singularity corresponds to the polynomial $P_n (x)  = x^n + v_c + \delta v$, dual on the string theory 
side to adding an $\cn = 2$ Liouville potential. This specific 
deformation of the superconformal \textsc{ad} fixed point breaks $U(1)_R$ to
$\zi_n$, matching precisely, as previously noted, the winding non-conservation
in the cigar $\slc$.

We have to stress  the fact that there is no decoupling between little string theory modes and 
gauge theory physics in the regime where the non-critical string dual is weakly coupled. The gauge 
theory, near the Argyres-Douglas point, has a low a low-energy scale of order\footnote{
For the deformation of the singularity that we consider, all the dyons have masses of the same 
order (at finite $n$).}  
$m_\textsc{dyon}$ and a high-energy scale $\Lambda$. The former is associated to the size of 
the cycles of the "small" torus, while the latter is associated to the cycles 
of the "large" torus of the Seiberg-Witten curve near the degeneration limit. The coupling 
constant of the gauge theory $\tau = \exp 2i\pi/n $ is independent of the separation of scales 
$\nicefrac{m_\textsc{dyon}}{\Lambda}$. In the string theory dual, the low energy 
scale is of the order of the D-brane mass. It is natural to associate the high energy scale 
$\Lambda$ with the (little) string scale $\nicefrac{1}{\sqrt{\alpha '}}$. At any rate the gauge 
theory scale cannot be much smaller since otherwise the singularities of the \textsc{sw} 
curve associated to the cycles of the large torus will be visible in the non-critical 
string dual. The coupling constant of the non-critical string dual is given by the 
ratio $g_\textsc{eff} \sim \nicefrac{1}{\sqrt{\alpha ' } m_\textsc{d}}$ of
those two scales. Then, is we want that the dyons masses are low compared to the high energy scale
$\Lambda$, the non-critical string dual is perturbative only for very low energies 
$E \ll m_\textsc{d} = \nicefrac{1}{\sqrt{\alpha '}g_\textsc{eff}}$.

\subsection{Chiral spectrum}
The chiral spectrum obtained from the string theory dual can be read from the analysis in 
section~\ref{nschir}. This special 
case of four-dimensional \textsc{lst} has been previously considered in~\cite{Giveon:1999zm}.
The bottom components of spacetime-chiral operators in the gauge theory 
coupling to the bulk operators discussed above are given by the following 
holographic dictionary  
\begin{equation}
\mathcal V^{(c,a)}_j = e^{-\varphi-\tilde \varphi}\ e^{ip_\mu X^\mu } \ e^{-Q \tilde \jmath \rho}
 \ e^{ i\frac{2 (j + 1) }{\sqrt{n (n + 2)}}\, (X_L  - X_R) } \ \ 
V_{j \, j+1 \, j +1}^{(2,2)} \ \qquad \longleftrightarrow  \qquad  \ 
s_{2j+2} (\phi)  \, ,
\end{equation}
where the symmetric polynomials appearing of the right-hand side are implicitly 
defined by eq.~(\ref{polchar}). However this correspondence is not exact, 
since there is some mixing between the operators appearing on the right-hand side. 
These effects have been computed in~\cite{Aharony:2004xn} in the context of six-dimensional 
\textsc{lst}. However the core of the computation had to do with the $SU(2)/U(1)$ part so 
they can be used also in the present context. The result is that the polynomials 
$s_r (\phi)$ have to be replaced by the following combination of multi-trace operators
\begin{equation}
s_{r} (\phi) \to \sum_{\ell =1}^{n} \sum_{r_i = 2}^r \
\frac{1}{\ell !} \left( \frac{1-r}{n} \right)^{\ell-1}
 \  \delta_{\sum r_i , r} \ 
\frac{1}{r_1} \mathrm{Tr}\, (\phi^{r_1} ) \cdots \frac{1}{r_\ell} \mathrm{Tr}\, (\phi^{r_\ell} ) \, .
\label{mixop}
\end{equation}
The two sets of operators agree only in the limit $\frac{r-1}{n} \to 1$. 
The R-charge in space-time of these chiral operators is given by the winding around the $U(1)$  as 
discussed previously:
\begin{equation}
R =  \frac{8 (j + 1) }{n + 2} \, .
\end{equation}
At the superconformal fixed point, the scaling dimension of these chiral primary fields 
can be obtained from the spacetime extended superconformal algebra as 
\begin{subequations}
\begin{align}
\Delta &= 2I + \frac{R}{2} \label{defdim2} \\ 
&= \frac{4 (j + 1) }{n + 2} \, ,
\end{align}
\end{subequations}
where, in~(\ref{defdim2}), $I$ is the spin under the $SU(2)_R$ symmetry which is zero for these 
operators. The allowed values of the spin $j$ in the  $SU(2)/U(1)$ coset are such that the string 
vertex operator obeys the Seiberg bound  
\begin{equation}
2j + 1 > \frac{n}{2} \, ,
\end{equation}
keeping half of the $n-1$ values of $2j +1 = 1, \ldots , n-1$ in the $\cn = 2$ minimal model, and 
then all the spacetime chiral primaries satisfy the unitarity bound $\Delta \geqslant 1$. 
The operator of highest 
dimension, i.e. for $2 j + 1 = n-1$, corresponds to the (non-normalizable branch of the) 
$\cn = 2$ Liouville potential. This string theory spectrum matches exactly the spectrum of relevant 
deformations of the superconformal fixed point predicted in the gauge theory~\cite{Argyres:1995jj}, 
as it has been already observed in~\cite{Giveon:1999zm}. We can extend straightforwardly 
the dictionary to the Ramond-Ramond sector, giving the correspondence
\begin{equation}
\mathcal{U}_j \ \longleftrightarrow  \ C_{\mu \nu} \mathrm{Tr} 
\left( F^{\mu  \nu} \phi^{2j-1} \right) \, ,
\end{equation} 
again up to multitrace corrections. 
The correlators of these off-shell observables lead to poles corresponding to
one-particle states created from the vacuum, 
see~\cite{Aharony:2004xn}. The poles from the \textsc{r-r} sector 
give $\lfloor \frac{n-1}{2} \rfloor$ 
Abelian gauge fields; the truncation on the number of 
$U(1)$ multiplets comes from the Seiberg bound. From the $\textsc{ns-ns}$ sector we get the same number of 
neutral scalars. Together with the \textsc{r-ns} and \textsc{ns-r} sectors 
they form $\lfloor \frac{n-1}{2} \rfloor$  gauge multiplets 
of $\mathcal{N}=2$ as expected from the gauge theory; indeed the dyons that are massless at the 
critical point are not charged under the other Abelian gauge multiplets, which decouple from the 
strongly interacting superconformal field theory.

%%%%%%%%%%%%%%%%%%%%%%%%%%%%%%%%%%%%%%%%%%%%%%%%%%%%%%%%%%%%%%%%%%
%%%%%%%%%%%%%%%%%%%D-branes%%%%%%%%%%%%%%%%%%%%%%%%%%%%%%%%%%%%%%%

\boldmath
\subsection{D-branes, dyons, and $\slc$ -- $\cn = 2$ Liouville duality}
\unboldmath

The light \textsc{bps} dyons of the gauge theory --~that would become massless at the Argyres-Douglas singularity~-- correspond 
in the doubly scaled little string theory to the localized D-branes of the cigar; similarly as we increase the 
string coupling constant at the tip (corresponding to approaching the superconformal fixed point) 
they become lighter. We will now consider the boundary worldsheet \textsc{cft} description of these D-branes, 
similar to the analysis carried out in six-dimensional \textsc{lst}~\cite{Israel:2005fn}.

Our goal is to find the couplings between the closed string modes and the D-branes, in other worlds 
to study the associated boundary states. The most non-trivial part of the computation comes from 
the boundary state associated to the localized B-brane in the super-coset $\slc$, which has been found only 
recently~\cite{Eguchi:2003ik,Israel:2004jt,Ribault:2003ss}. Then we need to combine it 
consistently with the contribution of an A-brane of the $\cn = 2$ minimal model $SU(2)/U(1)$ and of  
a D0-brane of the flat spacetime in the 
\textsc{gso}-projected theory (see~\cite{Eguchi:2003ik} for a very close analysis). It amounts to use 
standard orbifold techniques, and we end with D-branes characterized (apart from the fermionic labels 
which have to do with the orientation of the brane, and the position $\hat{\mathbf{y}}$ of the brane 
in flat space) 
by $(\hat \jmath , \hat m)$, corresponding respectively the $SU(2)$ spin and the $\zi_{2n}$ 
charge of a primary state of $SU(2)/U(1)$. Let's take a generic primary state $\mathcal{O}$ of the 
closed string theory, labeled by the left- and right- fermions numbers $(s_i,\bar s_i)=(a+2\upsilon_i, \bar a+2\bar\upsilon_i)$, 
the $SU(2)/U(1)$ labels $(j,m,\bar m)$, the linear dilaton imaginary  
momentum $-Q \tilde \jmath$ as well as the $U(1)$ charges of 
$(i\p X,i\pb X)$
\begin{equation}
(M,\bar M)= (2m+n s_4 + 4nr,-2\bar m +n \bar s_4+4n \bar r) \ \text{with} \quad 
m-s_4-4r = \bar m + \bar s_4 + 4 \bar r \mod 2(n+2) \, .
\end{equation}
Then the one-point function on the disc for a localized supersymmetric D-brane can be found to be, in the light-cone gauge  
\begin{eqnarray}
\langle \mathcal{O}^{(s_i ) ( \bar s_i)}_{p_\mu ;\, \tilde \jmath \, M\, \bar{M}  ; \, j\, m\, \bar m\ } 
(z, \bar z) \rangle_{\hat \jmath \, \hat m \, (\hat s_i) \hat{\mathbf{y}}}
 =  |z-\bar z|^{-\Delta -\bar \Delta} 
\sqrt{\frac{n+2}{2}} \frac{\nu^{\frac{1}{2}-\tilde \jmath}}{n}  \delta_{m,\bar m} \delta_{M,-\bar M} \ 
\delta_{s_1 , \bar s_1} \delta_{s_2 , -\bar s_2} \delta_{s_3 , \bar s_3}  \times
\nonumber \\  \times 
e^{i (p_2 \hat y^2 +  p_3 \hat y^3)}  e^{i \frac{\pi}{2} 
 \sum\limits_{\ell=1}^3  s_\ell \hat s_\ell}  \ 
e^{-i \pi \frac{m \hat m}{n}}\ \frac{\sin \pi \frac{(1+2j)(1+2\hat \jmath)}{n}}{\sqrt{\sin \pi \frac{1+2j}{n}}} 
\frac{\Gamma (\tilde \jmath + \frac{M}{2(n+2)}-\frac{s_2}{2})
\Gamma (\tilde \jmath+\frac{\bar M}{2(n+2)}-\frac{\bar s_2}{2})}{
\Gamma (2 \tilde \jmath-1) \Gamma (1+\frac{(n+2)(2 \tilde \jmath-1)}{2n})} \, , \nonumber \\
\label{onept}
\end{eqnarray}
with $\nu = \Gamma(1/2-1/n) /\Gamma (3/2+1/n)$. 
This expression has poles for values of $\tilde \jmath$ corresponding to discrete representations 
of the $\slr$ algebra. In particular the poles associated to the operators~(\ref{vca}) can be understood 
as the couplings between the vector multiplets and the dyon hypermultiplets dual to the D-branes in the gauge theory 
effective action. One can see that there is a one-to-one correspondence, see fig.~\ref{swbranes}, 
between the geometry of the D-brane 
in the coset $SU(2)/U(1)$ --~whose data $(\hat \jmath, \hat m)$ gives the endpoints of the 
A-type D-brane among the $n$ special points on the boundary of the disc~\cite{Maldacena:2001ky}\footnote{
To be precise, the D-brane is stretched between the angles   
$\pi (\hat m -2\hat \jmath-1)/n$ and $\pi (\hat m +2\hat \jmath+1)/n$.}~-- and cycles of the 
Seiberg-Witten curve whose periods give the masses of the corresponding dyons in the gauge theory 
(see also~\cite{Lerche:2000uy}). For the particular deformation of the \textsc{ad} singularity preserving the  
$\zi_n$ symmetry that we consider, any homology cycle encircling two branch points of the curve will 
give a stable \textsc{bps} dyon~\cite{Shapere:1999xr}. 
\FIGURE{ \epsfig{figure=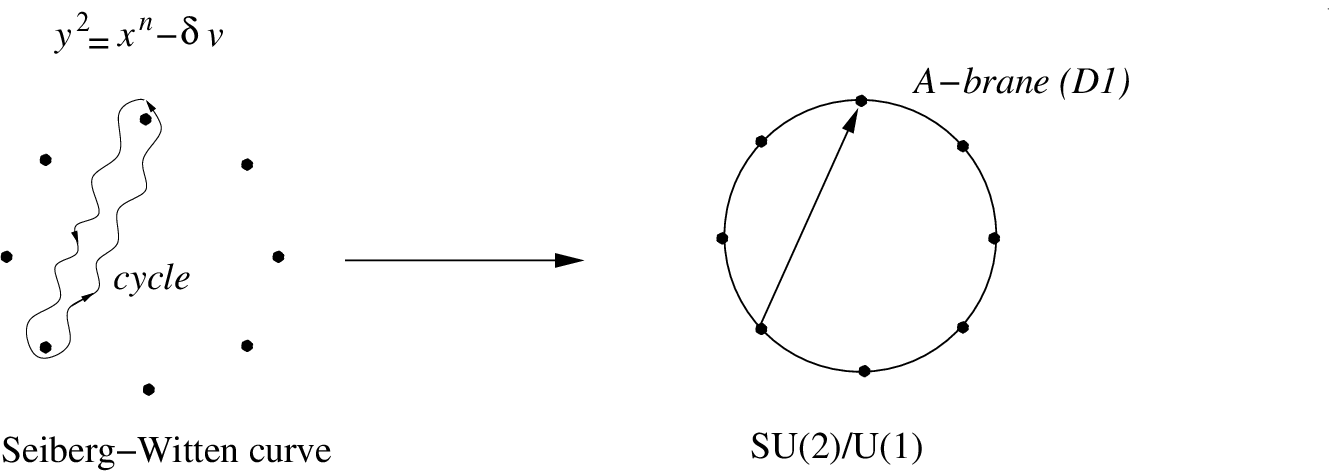, width=100mm} \caption{Periods of the \textsc{sw} curve and 
corresponding D-branes in $SU(2)/U(1)$.}
\label{swbranes}}
In the classical construction of the gauge theory with D4-branes stretched between NS5-branes, they 
are D2-branes with a disc topology ending on both NS5-branes and two D4-branes. In the 
strongly coupled quantum regime that is captured by the non-critical string there is no sharp bulk  
geometrical picture of those D-branes.

Using this one-point function one can first compute the annulus amplitude in the 
closed string channel, which is a sesquilinear form on the coefficients of the boundary states, with 
the insertion of the character for each corresponding representation propagating between the boundaries.  
Upon a modular transformation to the open string channel, using 
the modular transformation formulas given in the appendix, one obtains the following one-loop 
amplitude for the open strings stretched between any two of those D-branes:
\begin{eqnarray}
Z_{\textsc{open}} = \frac{q^{(\frac{\hat{\mathbf{y}}'-\hat{\mathbf{y}}}{2\pi})^2}}{
\eta^2} \sum_{\{ \upsilon_i\} \in (\zi_2)^3} 
\frac{1}{2} \sum_{a,b \in \zi_2} (-)^{a+b(1+\upsilon_1+\upsilon_2+\upsilon_3+m)} 
\frac{\Theta_{a+2\upsilon_1+\hat s_1 - \hat s_1 ',2}}{\eta} 
\ \times 
\nonumber \\ \times \ 
\sum_{2j = 0}^{n-2} \sum_{m \in \zi_{2n}} N^{j}_{\hat \jmath \hat \jmath '} C^{j\, (a+2\upsilon_3+\hat s_3 - 
\hat s_3 ')}_{2m+a+\hat m - \hat m'} \ 
Ch_{\mathbb{I}}^{(a+2\upsilon_2+\hat s_2 -\hat s_2 ')} (m)  \, .
\end{eqnarray}
First, local worldsheet supersymmetry imposes the condition 
$\hat s_\ell ' - \hat s_\ell = 0 \mod 2, \, \forall \ell$. 
Then this sector of open strings will be spacetime-supersymmetric provided that
\begin{equation}
\sum_{\ell=1}^{3} ( \hat s_\ell ' - \hat s_\ell )+ \frac{2(\hat m' - \hat m)}{n} = 0 \mod 4 \, .
\end{equation}
One can show that the symplectic form on the charge vectors of any pair of \textsc{bps} 
dyons is given by the open string Witten index for strings stretched between the corresponding 
D-branes~\cite{Eguchi:2003ik}.  
 
Open strings with both ends on any on these D-branes will always give a supersymmetric 
spectrum. The massless open string modes correspond to the dimensional reduction of a four-dimensional 
$\cn = 1$ gauge multiplet, for example in the \textsc{ns} sector we have the bosonic states 
\begin{equation}
(\xi^2 \pm i\xi^3) |0\rangle_\textsc{ns} \otimes |j=0,m=0,s=0 
\rangle_\textsc{su(2)/u(1)} \otimes | r= 0 \rangle_\textsc{sl(2,r)/u(1)} \, .
\label{gaugefopen}
\end{equation}
Below the string scale we have also "light" multiplets of mass $\alpha' m^2 = \nicefrac{j(j+1)}{n}$ for 
all the non-zero values of $j$ allowed by the fusion rules $N^{j}_{\hat \jmath \hat \jmath '}$. 
Thus the dynamics of the massless degrees of freedom for $N$ coincident D-branes 
is the quantum mechanics of a supersymmetric gauged matrix model. 
Upon T-duality in the flat space-like directions $x^{1,2,3}$, 
one can construct pure $\cn = 1$ \textsc{sym} in four dimensions, generalizing the conifold case studied 
in~\cite{Fotopoulos:2005cn,Ashok:2005py}. 

The mass of the D-branes can be obtained from the one-point function of a graviton vertex 
operator, coming from the continuous representations of $\slr$ and the identity representation of $SU(2)$:
\begin{equation}
h^{\mu \nu} = e^{-\phi-\tilde \phi} e^{i p_\mu X^\mu} \psi^{( \mu} \tilde \psi^{\nu )} 
e^{-(\frac{Q}{2}+ i Q\, P)\rho} V^{(0,0)}_{0\, 0\, 0} \, .
\end{equation}
To compute the massless tadpole associated to the graviton, we need to compute the overlap between 
the boundary state and the graviton closed string state 
with the insertion of a closed string propagator:\footnote{For a similar computation for the conifold 
see~\cite{Ashok:2005py}}
\begin{eqnarray}
H^{\mu \nu} (p_\mu, \, P) &=& \langle h^{\mu \nu} | \frac{\alpha'}{4\pi}
 \int_{|z|\leqslant 1} \frac{\di^2 z}{z \bar z} \ 
z^{L_0 -\frac{1}{2}} \ 
\bar{z}^{\bar L_0 - \nicefrac{1}{2}} 
\ | B \rangle_{\hat \jmath} \nonumber \\ & = & \frac{1}{p_\mu p^{\mu} +\frac{2}{\alpha ' } Q^2 (P^2 
+\nicefrac{1}{4}) }  \ 
\langle h^{\mu \nu} \rangle_{\hat \jmath}^\textsc{disc} \, .
\end{eqnarray}
We have suppressed the extra label $\hat m$ of the brane since it
corresponds to the position of a D-brane of a given mass and doesn't change the result. 
Then after a Fourier transform to position space we would obtain that the mass of the D-brane scales like
\begin{equation}
 m_\textsc{d} \sim \frac{1}{\sqrt{\alpha'}\, g_{\textsc{eff}}} \ 
\sin  \frac{\pi (1+2\hat \jmath)}{n} \, ,
\label{massdbranews}
\end{equation}
in terms of the effective string coupling of the cigar~(\ref{cigcoupl}) weighting this 
disc amplitude. This computation is sensitive to the proper normalization of the fields, 
and it would be quite difficult to give a precise definition of the \textsc{adm} mass 
in this strongly curved background. The normalization that we would find depends on $n$, but 
since we are not interested in the large $n$ limit in this paper it won't affect the 
discussion. However the ratio of masses of various dyons are not sensitive to this, 
and are precise predictions of the string dual of the gauge theory.

\boldmath
\paragraph{Masses of hypermultiplets and the cigar / $\cn = 2$ Liouville  duality}
\unboldmath
We would like now to compute in the gauge theory the masses of the light hypermultiplets 
corresponding to the localized D-branes. As discussed in the previous section 
the gauge theory is near a superconformal fixed point in the 
moduli space, for which the \textsc{sw} curve degenerates to a "small" torus of genus $
\lfloor \frac{n-1}{2} \rfloor$ described by the curve 
\begin{equation}
y^2 \simeq \pm \Lambda^n (x^n -\delta v ) \, .
\end{equation}
Then the branch points of the \textsc{sw} 
curve are distributed evenly as $x_\ell = (\delta v)^{1/n} e^{2i\pi \ell/n}$, and the 
masses of the dyons wrapping cycles of this Riemann surface are of the same order 
for finite $n$. The exact expression for the masses will be 
given by integrating over these cycles the Seiberg-Witten one-form
\begin{equation}
\lambda = \frac{1}{2i\pi} \frac{\partial P_n (x)}{\partial x} \frac{x\, \di x}{y} \, , 
\end{equation}
around any pair of branch points. If we choose the parameterization $\delta v = 2 \varepsilon^n$, and perform the 
rescaling $x=\varepsilon z$, the \textsc{sw} differential is given 
in the vicinity of the \textsc{ad} point by 
\begin{equation}
\lambda \simeq  \frac{ \varepsilon^{n/2+1}}{\Lambda^{n/2}} \ \frac{n}{2i\pi} 
\frac{z^n \, \di z}{\sqrt{z^n-2}} \, .
\end{equation}
Such that all the masses of the \textsc{bps} dyons near the \textsc{ad} point are given by 
integrating the Seiberg-Witten one-form over the corresponding one-cycles of the 
torus (rescaled by $\varepsilon$)
\begin{equation}
m_\textsc{dyon}^{(i)} \simeq \frac{|\varepsilon|^{n/2+1}}{\Lambda^{n/2}} 
\left| \sqrt{2}\, \frac{n}{2i\pi} \int_{C_{(i)}} 
\frac{z^n \, \di z}{\sqrt{z^n-2}} \right| \, .
\label{massdyons}
\end{equation}
The precise expression of those masses are given in terms of the multiple hypergeometric 
functions~\cite{Akerblom:2004cg} but is not necessary for our purposes. 
Since, for finite $n$, all these dimensionless integrals are of the same order 
(and don't depend neither on $\Lambda$ nor on $\varepsilon$) we
get the scaling relation for the dyons masses $m_\textsc{dyon}$ in terms of $\delta v$ as: 
\begin{equation}
m_\textsc{dyon}^2 \sim (\delta v)^{\frac{n+2}{n}} \, . 
\label{scmdyons} 
\end{equation}

It is very interesting that this non-trivial scaling relation can be 
found from worldsheet non-perturbative effects in the dual string theory. 
On the one hand, according to the holographic dictionary discussed above,  
$\delta v$, the vacuum expectation value of $s_n (\phi) = (-)^n \det (\phi)$ away 
from the critical value at fixed point, 
corresponds to the coupling constant of the $\cn = 2$ Liouville potential~(\ref{n2pot}). 
On the other hand, the coefficient of cigar perturbation~(\ref{cigpert}) is 
related to the effective string coupling constant. The coupling 
constant in type \textsc{iia} superstrings is quite generically related to the masses of the localized 
supersymmetric D-branes, 
corresponding to the non-perturbative \textsc{bps} states of the theory, 
as $\alpha ' m_{d}^2 \sim \nicefrac{1}{g_{\textsc{eff}}^2}$. 
In our particular example, the masses are given by~(\ref{massdbranews}).\footnote{
As we have several masses in the problem, corresponding to the various D-branes in 
the coset $SU(2)/U(1)$,  
we identify the effective coupling constant with the inverse of the common mass for the 
elementary cycles (giving the dyons of lowest mass).} 
Putting everything together, 
the scaling relation for the dyons masses predicted by the gauge theory, 
see eq.~(\ref{scmdyons}), gives a scaling relation between the coefficient 
of the geometric cigar and $\cn = 2$ Liouville perturbations on the 
worldsheet.

The scaling relation that we obtain 
matches exactly the worldsheet prediction from the duality between 
the supersymmetric cigar $\slc$ at level $k$ and $\cn = 2$ Liouville, see~\cite{Giveon:2001up}:
\begin{equation}
\mu_\textsc{Liouville} = - \frac{k}{2\pi} \left(  \frac{ \Gamma (1+\nicefrac{1}{k} )}{
\Gamma (1-\nicefrac{1}{k})} 
\ k  \pi \, \mu_\textsc{cigar} \right)^{k/2} \, ,
\label{liouvcigdual}
\end{equation}
with $k = \nicefrac{2n}{n+2}$ in the present case. The essential meaning of this relation is that the cigar 
\textsc{cft}, i.e. the super-coset theory $\slc$, receives non-perturbative worldsheet 
corrections in the form of a winding condensate, the $\cn = 2$ Liouville interaction, with 
a coefficient given by~(\ref{liouvcigdual}) (the exponent itself follows from \textsc{kpz} 
scaling~\cite{Knizhnik:1988ak}).
Both types of operators are needed 
as screening charges for perturbative computations (in the worldsheet sense) 
of the \textsc{cft} correlators. It is very interesting to see that these non-perturbative effects 
are mapped through the holographic duality to the exact Seiberg-Witten solution  of the gauge theory, 
i.e. the masses of the dyons as functions of the gauge-invariant coordinates on the moduli 
space of the Coulomb branch. We expect that the precise coefficient appearing in this relation 
could be matched to the gauge theory prediction if we knew how to compute the normalization of the D-brane 
masses properly. We expect also that the ratio of the masses for dyons corresponding to cycles of different 
lengths, using~(\ref{massdyons}), reproduce the results from the string theory dual 
using~(\ref{massdbranews}) since this quantities are independent of the double scaling 
parameter.

%%%%%%%%%%%%%%%%%%%%%%%%%%%%%%%%%%%%%%%%%%%%%%%%%%%%%%%%%%%%%%%%%%%%%%%%%%%%%%%%%
%%%%%%%%%%%%%%%%%%%%%%%%%%%%%%%%%%%%%%%%%%%%%%%%%%%%%%%%%%%%%%%%%%%%%%%%%%%%%%%%%
%%%%%%%%%%%%%%%%%%%N=1 MODELS%%%%%%%%%%%%%%%%%%%%%%%%%%%%%%%%%%%%%%%%%%%%%%%%%%%%
%%%%%%%%%%%%%%%%%%%%%%%%%%%%%%%%%%%%%%%%%%%%%%%%%%%%%%%%%%%%%%%%%%%%%%%%%%%%%%%%%
\boldmath
\section{Non-critical superstrings with $\cn =1$ supersymmetry}
\unboldmath
\label{n1lst}
Starting from the previous construction of strings duals of $\cn = 2$ gauge theories,  
it is possible to construct new four-dimensional non-critical superstring 
theories with only one spacetime supersymmetry, that will be solvable string duals of 
$\cn = 1$ gauge theories and their  little string theory \textsc{uv} completion. 
These theories will also be at non-trivial superconformal fixed points, or in the neighborhood of those, 
and the spectrum of massless string modes will give the scaling dimensions of chiral 
operators in the superconformal field theory. 

\subsection{Asymmetric orbifolds of four-dimensional non-critical superstrings}
To construct these new non-critical strings we will perform an asymmetric orbifold of the worldsheet 
\textsc{cft}, quite similar to a lens space, acting only on the left-movers of the worldsheet. 
In the context of six-dimensional \textsc{lst}\footnote{
Whose holographic dual is the \textsc{chs} background~\cite{Callan:1991dj}
$\mathbb{R}^{5,1}\times \mathbb{R}_Q \times SU(2)$.}, one can 
replace the $SU(2)_k \sim [ \left. SU(2)/U(1) \right|_k \times U(1)_k ]/\zi_k$ 
super-\textsc{wzw} model by a lens space, i.e. a $\sfrac{\zi_p}{SU(2)_k}$ chiral 
orbifold~\cite{Giddings:1993wn} acting on the $SU(2)$ group elements as $g \to \exp{\frac{2i\pi}{p} \sigma_3} \, g$. 
The six-dimensional holography associated to this lens space 
has been studied in~\cite{Diaconescu:1998pj}.  

In the four dimensional little string theory studied in this paper we can  construct a similar 
asymmetric orbifold structure, this time acting on $[\left. SU(2)/U(1) \right|_n \times U(1)_{2n(n+2)}]$, provided 
that we choose the level of the coset as  
\begin{equation}
n=pp' \quad  \text{with} \qquad p,\, p' \in \zi \, .
\end{equation}
This restriction will be understood later from the gauge theory point of view. 
The partition function~(\ref{partcontred}) of the original model for the continuous representations is replaced by 
the asymmetric $\zi_p$ orbifold partition function 
\begin{eqnarray}
Z_\textsc{cont} (\tau , \bar \tau ) 
=  \frac{1}{4\pi^2 \alpha' \tau_2} 
\frac{1}{\eta^2 \bar \eta^2}\ \frac{1}{4} \sum_{a,b,\bar a , \bar b \in \zi_2} 
\sum_{\{ \upsilon_\ell \} , \{ \bar \upsilon_\ell \} \in (\zi_2)^4}
(-)^{a+\bar a + b (1 + \sum_i \upsilon_i) + 
\bar b (1 + \sum_i \bar \upsilon_i)+\bar a \bar b }\times \nonumber \\ \times
\frac{\Theta_{a+2\upsilon_1,2} \Theta_{\bar a+2\bar \upsilon_1,2}}{\eta \bar \eta}\ 
\int_0^\infty \di P  \sum_{2j =0}^{n-2}  \sum_{m , \bar m \in \zi_{2n}} 
\sum_{ r \in \zi_{n + 2} } \ 
\delta_{m -a-2\upsilon_4-4 r ,\bar m +\bar a+2\bar \upsilon_4 +4 \bar r   \mod 2(n + 2)}\times 
\nonumber \\  
\nonumber \\ \times \ 
Ch_c^{(a+2\upsilon_2)} (P,2 m + n (a+2\upsilon_4)+ 4 n r )
\bar{Ch}_c^{(\bar a+2\bar \upsilon_2)} (P,-2 \bar m + n (a+2\bar \upsilon_4)+ 4 n \bar r )\
\vphantom{\frac{1}{2_{2_2}}} \times \nonumber \\
\times \ 
\sum_{\gamma \in \zi_{p}}  \ \delta_{m-p' \gamma,0 \mod p}  \ \ 
C^{j\, (a+2\upsilon_3)}_{m-2p' \gamma} \  \bar C^{j\, (\bar a+2\bar \upsilon_3)}_{\bar m} \, ,\nonumber \\
\label{partcontredorb}
\end{eqnarray}
that is modular invariant as can be checked explicitly.  
The states invariant under the $\zi_p$ orbifold action are selected by the Kronecker delta in the last line and 
the twisted sectors, requested by modular invariance, are labeled by the integer $\gamma$ taking values 
in $\zi_p$. In particular the spacetime supercharges  from the left-movers in the 3+1 \textsc{lst}, 
given by eq.~(\ref{stsupercharges}), will be projected out because they have $m=\pm 1$. On the other 
hand, since the orbifold doesn't act on the right, the spacetime supercharges from the 
right-movers are preserved. We end up with a non-critical string theory with only $\cn = 1$ 
supersymmetry in four dimensions, albeit without Ramond-Ramond fluxes. 
From this partition function we can read the spectrum of string states that are dual to off-shell 
operators in the dual theory. Our aim will be to show that this non-critical string theory provides an example of holographic dual 
of $\cn = 1$ gauge theories that is exactly solvable. This theory will also flow to a non-trivial superconformal 
fixed point, if there is no potential along the linear dilaton direction.

\subsection{Chiral spectrum and double scaling limit}
Let's first discuss the spacetime chiral operators that we can read from this partition function. 
The $U(1)_{\tilde R}$ symmetry that appears in the $SU(2,2 |1)$ superconformal algebra 
gives the scaling dimension (at the superconformal fixed point) of the chiral/antichiral primary operators as
\begin{equation}
\Delta \ = \ \frac{3}{2} |\tilde R |\  \, . 
\label{scalspecorb}
\end{equation}
This $U(1)_{\tilde R}$ is given in terms of the $\cn = 2$ $U(1)_\textsc{R}$ charge $R$, see eq.~(\ref{rch}), 
and  the $U(1) \subset SU(2)_\textsc{r}$ charge $m_\textsc{r}$, which are both R-symmetries of the $\cn = 1$ theory, 
by the following linear combination 
\begin{equation}
\tilde{R} = \frac{R-4m_\textsc{r}}{3} = \frac{2}{3Q} \oint \left[ \p X - \pb X - 2 (\p X + \pb X) \right] \, ,
\label{rchargen1}
\end{equation}
such that the surviving supercharges have charge $\pm 1$ (they have only right-moving momenta from the 
worldsheet point of view).\footnote{If we would have done instead the orbifold on the right-moving worldsheet 
\textsc{cft}, preserving the left-moving spacetime supercharges, the correct definition would have been 
$\tilde R = \frac{R+4m_\textsc{r}}{3}$.}
Let us look first at the untwisted \textsc{ns-ns} sector. We can build operators similar to those  
$\cn = 2$ models, see eq.~(\ref{vca}), but they have to be invariant under the 
orbifold action:
\begin{equation}
\mathcal{V}_{\frac{pN}{2}-1}^\textsc{u} = 
e^{-\varphi-\tilde \varphi} e^{-Q \tilde \jmath \rho} 
e^{i \sqrt{\frac{p}{p' (pp'+2)}} N (X_L - X_R)} 
V_{\frac{pN}{2}-1 ,\, \frac{pN}{2}, \, \frac{pN}{2}}^{(2,2)}    
\end{equation}
constructed from $SU(2)$ representations of spin $j = pN/2-1$. 
These operators are dual to chiral operators in the $\cn = 1$ gauge theory which follows 
from the fact that they were the bottom components of the vector multiplet in the $\cn = 2$ theory that 
was orbifoldized. Using the definition~(\ref{rchargen1}) of the R-charge in spacetime, we obtain the following spectrum of scaling dimensions for these operators at the superconformal fixed point:
\begin{equation}
\Delta_\textsc{u} = \frac{4pN}{pp'+2} \quad \text{for} \qquad N > \frac{p'}{2} + \frac{1}{p} \, .
\label{untwchir}
\end{equation}
The scaling dimension of these chiral operators is the same as those of the 
$\cn = 2$ chiral operators of the parent theory they are coming from.\footnote{This could be 
understood e.g. in the M-theory AdS description of the superconformal theory 
discussed above, since of course the 
asymptotic behavior of the bulk fields in $AdS_5$ coupling to these chiral operators won't change 
with the orbifold;  it will simply project out eigenfunctions for the Laplacian on $X^6$ 
that are not invariant under the orbifold action.}

The orbifold theory contains also twisted sectors, labeled by $\gamma \in \zi_{p}$. For each such 
sector we can construct chiral operators in spacetime as follows:
\begin{equation}
\mathcal{V}_{\frac{p\gamma}{2}-1}^\textsc{t} = 
e^{-\varphi-\tilde \varphi} e^{-Q \tilde \jmath \rho} 
e^{i \sqrt{\frac{p'}{p (pp'+2)}} \gamma (X_L - X_R)} 
V_{\frac{p'\gamma}{2} -1 , \, -\frac{p'\gamma }{2} ,\, \frac{p' \gamma}{2}}^{(2,2)}  
\end{equation}
These states violate the integrality condition of the left worldsheet $\cn = 2$ R-charge because 
the $\zi_{2n}$ charge for the left-moving sector of $SU(2)/U(1)$ has an 
opposite sign w.r.t. the $\slc$ contribution,  
breaking the spacetime supersymmetry associated to the $\cn = 2 $ spectral flow. 
However the spacetime supersymmetry associated with the right-movers 
is preserved. Again, eq.~(\ref{rchargen1}) gives the R-charge and thus the scaling dimension in spacetime 
of those operators:
\begin{equation}
\Delta_\textsc{t} = \frac{4p' \gamma }{pp'+2} \quad \text{for} \qquad \gamma > \frac{p}{2} + \frac{1}{p'} \, .
\label{twchir}
\end{equation}
It is quite interesting to note that the T-duality between the 
$\zi_p$ and the $\zi_{p'}$ orbifolds, similar to the T-duality between the lens spaces 
$
\sfrac{\zi_p}{SU(2)}_{pp'} \ \stackrel{\textsc{T}}{\longleftrightarrow}\ 
\sfrac{\zi_{p'}}{SU(2)}_{pp'}\, ,
$
translates into a symmetry in the spectrum for the gauge theory in the neighborhood 
of the superconformal fixed point. Indeed the 
chiral operators from the untwisted sector~(\ref{untwchir}) and from the twisted sectors~(\ref{twchir}) 
are exchanged under $p \leftrightarrow p'$, giving overall the same spectrum of scaling dimensions 
in the four-dimensional superconformal field theory. 

As in the $\cn = 2$ models, it is possible to regularize the strong coupling region 
of the linear dilaton by taking a double scaling limit, i.e. by 
replacing the $\mathbb{R}_Q \times U(1)$ factor by a super-coset $\slc$. 
The structure of the coset 
\textsc{cft} is not affected by the orbifold. In particular we see that the 
$\cn = 2$ Liouville operator is still in the spectrum, and can be viewed either 
as an untwisted sector operator~(\ref{untwchir}) with $N=p'$ or a twisted 
sector one~(\ref{twchir}) with $\gamma = p$. 
Then the one-loop vacuum amplitude will also contain a 
contribution for discrete representations, that will be similar to~(\ref{partdisc}) upon replacing 
\begin{equation}
 C^{j\, (a+2\upsilon_3)}_{m}  \quad \to \quad 
\sum_{\gamma \in \zi_{p}}  \ \delta_{m-p' \gamma,0 \mod p} \ \ 
 C^{j\, (a+2\upsilon_3)}_{m-2p' \gamma}\, ,
\end{equation}
as for the continuous representations~(\ref{partcontredorb}). The worldsheet chiral operators 
belonging to this discrete spectrum will be in one-to-one correspondence with relevant deformations 
of the $\cn =1$ superconformal field theory in spacetime, as in the $\cn = 2$ models.  

\subsection{Brane setup for the gauge theory}
\label{branesetup}
In order to find the gauge theory duals of these $\cn = 1$ non-critical strings 
in the semi-classical regime, we can use the NS5/D4 construction 
in type \textsc{iia} superstrings  of the $\cn =2$ \textsc{sym} theories discussed 
in section~\ref{N2dual}, and show how the asymmetric orbifold of the non-critical string is implemented 
at the level of this brane setup.  
We consider again a pair of NS5-branes of worldvolume $x^{0,1,2,3,4,5}$ located at $x^6=0,\ x^6=L$ and 
$x^7=x^8=x^9=0$. They preserve the supercharges corresponding to spinors solutions of 
\begin{equation}
\eta_{\, ^\textsc{l}_\textsc{r} } = \pm \gamma^0 \gamma^1 \gamma^2 \gamma^3 \gamma^4 \gamma^5 \ 
\eta_{\, ^\textsc{l}_\textsc{r} } \, ,
\end{equation}
where the left and right supercharges $\eta_\textsc{l}$ and $\eta_\textsc{r}$ have opposite chiralities. 
We stretch between the fivebranes $pp'$ D4-branes, of worldvolume $x^{0,1,2,3,6}$, 
distributed in the $x^{4,5}$ plane. They preserve the following supercharges 
\begin{equation}
\eta_\textsc{l} = \gamma^0 \gamma^1 \gamma^2 \gamma^3  \gamma^6 \ \eta_\textsc{r} \, .
\end{equation}
On top of this brane configuration we can add a $\mathbb{C} ^2/ \zi_p$ orbifold, acting in the planes 
$x = x^4 + ix^5$ and $h=x^8 + ix^9$ as follows 
\begin{equation}
x \to e^{-\frac{2i\pi}{p}}\ x \quad , \qquad h \to e^{\frac{2i\pi}{p}}\  h \, ,
\label{orbsup}
\end{equation}
which is a symmetry of the brane configuration, provided that the D4-branes are distributed in a \mbox{$\zi_p$-symmetric} fashion in the $(4,5)$ plane; in other words we consider $p'$ D4-branes in the fundamental domain of the orbifold. 
This orbifold preserves the left and right supercharges that satisfy 
\begin{equation}
\eta_{\, ^\textsc{l}_\textsc{r} } =  \gamma^4 \gamma^5 \gamma^8 \gamma^9  \ 
\eta_{\, ^\textsc{l}_\textsc{r} }  \, ,
\end{equation}
leading overall to $\cn =1$ in four dimensions. 
Such models have already been considered in~\cite{Lykken:1997gy}, however the construction of the dual 
non-critical string is new. 

The identification of this brane construction --~that describes the gauge theory dual in the 
semi-classical regime~-- with our $\cn = 1$ four-dimensional non-critical string 
is as follows. The action of the orbifold on the $x=x^4+ix^5$-plane 
can be seen from the dual formulation of the four-dimensional non-critical string 
as a singular Calabi-Yau three-fold compactification in type \textsc{iib}, 
see e.g.~\cite{Giveon:1999zm}. 
One can relate the $\cn = 2$ non-critical strings of the sort~(\ref{back4lstred}) to the decoupling 
limit of the singular \textsc{cy}$_3$ given by the hypersurface $x^n + y^2 + uv = 0$ 
embedded in $\mathbb{C}^{4}$.\footnote{This description 
is indeed related by a T-duality to the description in terms of an NS5-brane wrapping 
the curve $x^n+y^2=0$ discussed above.} 
To make the connection more obvious 
we consider another representation of the worldsheet \textsc{cft} as the 
infrared limit of a Landau-Ginzburg model with the potential 
$X^{pp'} + Y^2 + UV+\mu Z^{-\frac{2n}{n+2}}$, the last factor corresponding to the $\cn = 2$ 
Liouville/cigar theory if the singularity is resolved.  
The elements of the chiral ring of this theory are the monomials 
$X^{2j+2}$, dressed appropriately by the $\cn = 2$ Liouville system 
as we explained above in a different language. 
The orbifold that we consider in the non-critical string~(\ref{partcontredorb}) 
leaves invariant only the elements of
the chiral ring of the form $X^{pN}$, $N \in \mathbb{N}$. It is then 
identified as a $\zi_p$ rotation in the complex $x=x^4+ix^5$ plane, in the brane 
construction described above.  
The action in the $h$-plane, see eq.(\ref{orbsup}), is dictated by supersymmetry (because a
$\mathbb{C}/ \zi_p$ orbifold alone breaks supersymmetry). The symmetries of the 
brane construction allows only this orbifold to act on a $U(1)$ subgroup of the $SU(2)_R$ symmetry 
corresponding to rotations in the $x^{7,8,9}$ overall transverse directions,  
preserving $\cn = 1$ supersymmetry in four dimensions.

Let us analyze briefly the gauge theory corresponding to this configuration. 
We start with $pp'$ D4-branes having on their worldvolume an $\cn = 2$ $U(pp')$ gauge multiplet in 4+1 dimensions. 
It contains an $\cn =1$ gauge multiplet, whose scalar component corresponds to the fluctuations along $x^7$, and a  
hypermultiplet, whose four scalar components correspond to fluctuations along
$x^{4,5,8,9}$. 
The action of the orbifold on the $pp'$ D4-branes gives 
an $U(p')^{p}$ quiver gauge theory with $\cn = 1$ in five dimensions. 
Now we suspend these D4-branes between two NS5-branes along the
$x^6$ direction; since the NS5-branes 
are very heavy we consider only the degrees of freedom living on the D4-branes.
Because the D4-branes end on NS5-branes, the boundary conditions at their endpoints 
will remove part of the massless fields and  break half of 
the supersymmetry on their worldvolume, as suggested in~\cite{Hanany:1996ie,Witten:1997sc}. 
These boundary conditions will indeed 
set to zero the fluctuations along $x^{6,7,8,9}$. We end up at low energies 
(compared to the inverse of the distance between the fivebranes) with $\cn = 1$ gauge multiplets 
in four dimensions and bifundamental chiral multiplets. 
As for the $\cn = 2$ models the diagonal $U(1)$ factor is frozen. The other 
$U(1)$'s are also all anomalous (for $p>2$), leaving an $SU(p')^p$ quiver~\cite{Lykken:1997gy}. 
For a generic vacua  this gauge symmetry is broken to $U(1)^{p' -1}$, thus we are in 
an Abelian Coulomb phase. 
We will give below a better derivation of the open string theory 
living on the D4-branes.

%%%%%%%%%%%%%%%%%%%%%%%%%%%%%%%%%%%%%%%%%%%%%%%%%%%%%%%%%%%%%%%
%%%%%%%%%%%WORLDSHEET CONSTRUCTION%%%%%%%%%%%%%%%%%%%%%%%%%%%%%
%%%%%%%%%%%%%%%%%%%%%%%%%%%%%%%%%%%%%%%%%%%%%%%%%%%%%%%%%%%%%%%

\subsection{Boundary worldsheet CFT construction}
The construction that we outlined in the previous section gave the correct field 
content of the gauge theory on the D4-branes, however it is somehow heuristic since the analysis 
of the D4-branes ending on NS5-branes was only qualitative. 
It is possible to give a precise description of this configuration, including 
the backreaction of the NS5-branes, along the lines 
of~\cite{Israel:2005fn} where the $\cn = 2$ case was studied.  
We analyze the system from the worldsheet point of view, starting from the 
worldsheet \textsc{cft} describing the background for the near-horizon geometry 
of two separated NS5-branes in type \textsc{iia}. 
This is a good approximation since the NS5-branes are considered in these constructions are very heavy, non-fluctuating 
objects. Our aim is to derive more rigorously the construction of the $\cn = 1$ gauge theory 
explained above, by constructing the D4-brane boundary state in the presence of the fivebranes and of the orbifold.

The worldsheet \textsc{cft} for two flat parallel fivebranes corresponds to
the string background 
$\mathbb{R}^{5,1} \times \mathbb{R}_{Q=1} \times SU(2)_2$, i.e. eight-dimensional non-critical superstrings, which 
is a particular case of the \textsc{chs} background.  The two fivebranes 
can be separated in the transverse space if we replace $\mathbb{R}_Q \times U(1)_2$ with the super-coset 
$\slc$ at level $k=2$. The supersymmetric SU(2) \textsc{wzw} model at level 
two contains only the fermionic $SU(2)_2$ affine currents (since the 
bosonic algebra is at level zero), constructed with three fermions $\psi^3, \, \psi^{\pm}=(\psi^1 \pm i\psi^2)/\sqrt{2}$:
\begin{equation}
J^3 = \psi^+ \psi^- \quad , \qquad J^{\pm} = i\sqrt{2} \, \psi^\pm \psi^3 \, .
\end{equation}
In terms of those, and the remaining fermions $\xi^{2,3,4,5,\rho}$ of $\mathbb{R}^4 \times \mathbb{R}_Q$ 
the generators of the worldsheet $\cn = 2$ superconformal algebra --~which 
is actually enhanced to $\cn =4$~\cite{Kounnas:1990ud}~-- read 
\begin{eqnarray}
-2 iG^{\pm} &=& (\p X^2 \mp i\p X^3)(\xi^2 \pm i\xi^3 )+
(\p X^4 \mp i\p X^5)(\xi^4 \pm i\xi^5 ) \nonumber \\ && \qquad \qquad + \ 
\left[ \p \rho \pm i(\psi^1 \psi^2 -
 \xi^\rho \psi^3) \right] (\xi^\rho \pm i\psi^3) \nonumber \\
-iJ_R &=& \xi^2 \xi^3 + \xi^4 \xi^5 + \xi^\rho \psi^3 + \psi^1 \psi^2 
\end{eqnarray}
The action of the orbifold (\ref{orbsup}) in the transverse 
space translates into the following action on the fermionic $SU(2)$ left- and right-moving affine currents and thus on the free 
worldsheet fermions\footnote{This orbifold is different from the lens space since it acts non-chirally.}
\begin{eqnarray*}
J_3 & \to & J_3 \ , \ \ J^{\pm} \to e^{\pm \frac{2i\pi}{p}} J^{\pm} \quad \implies \qquad \psi^3 \to \psi^3 \ , \ \ 
\psi^\pm \to e^{\pm \frac{2i\pi}{p}} \psi^\pm  \nonumber \\
\tilde  J_3 & \to & \tilde  J_3 \ , \ \ \tilde J^{\pm} \to e^{\pm \frac{2i\pi}{p}} \tilde J^{\pm} 
\quad \implies \qquad \tilde \psi^3 \to \bar \psi^3 \ , \ \ 
\tilde \psi^\pm \to e^{\pm \frac{2i\pi}{p}} \tilde \psi^\pm \, ,
\end{eqnarray*}
together with an action of the orbifold 
on the free fields corresponding to the $(4,5)$ plane as an ordinary 
$\ci / \zi_p$ orbifold. This preserves of course the $\cn = 2$ algebra 
on the worldsheet but keeps only half of the spacetime supercharges. Next we consider the double 
scaling limit of this little string theory by separating the NS5-branes in the $(6,7)$ plane,  
which is not affected by the orbifold. The type \textsc{iia} 
partition function of the orbifold theory, for the continuous representations 
of $\slc$,  is given by 
\begin{eqnarray}
Z =   \frac{1}{4\pi^2 \alpha' \tau_2} \frac{1}{\eta^2 \bar \eta^2} \frac{1}{4p} \sum_{a,b,\bar a , \bar b } (-)^{a+b+\bar a + \bar b + \bar a \bar b}\ 
\frac{\vartheta \oao{a}{b}}{\eta}
\frac{\bar \vartheta \oao{\bar a}{\bar b}}{\bar \eta} 
\ \int \di P   \ 
\frac{(q \bar{q})^{\frac{P^2}{2}}}{\eta \bar \eta}   \frac{\vartheta \oao{a}{b}}{\eta}
\frac{\bar \vartheta \oao{\bar a}{\bar b}}{\bar \eta}\ \times  \nonumber \\ 
\left[ \frac{1}{4\pi^2 \alpha' \tau_2  |\eta|^4}\  \frac{\vartheta \oao{a}{b}^2}{\eta^2}
\frac{\bar \vartheta \oao{\bar a}{\bar b}^2}{\bar \eta^2}\ 
 + \!\!\!\!\!\!\!\!\! \sum_{\gamma, \delta \in \zi_p \neq (0,0)}
\frac{|\eta|^2}{\left| \vartheta \oao{1+2\gamma/p}{1+2\delta/p} \right|^2}  
\frac{\vartheta \oao{a-2\gamma/p}{b-2\delta/p}}{\eta}
\frac{\bar \vartheta \oao{\bar a - 2\gamma /p}{\bar b-2\delta /p}}{\bar \eta} 
\frac{\vartheta \oao{a+2\gamma/p}{b+2\delta/p}}{\eta}
\frac{\bar \vartheta \oao{\bar a + 2\gamma /p}{\bar b+2\delta /p}}{\bar \eta} 
\right]\nonumber \\
\end{eqnarray}
which has two supercharges in four dimensions, one from the left-movers and one from the right movers. 
The last $\vartheta \oao{\cdot}{\cdot} 
\bar \vartheta \oao{\cdot}{\cdot}/\eta \bar \eta$ factor corresponds to the fermions $(\psi^{\pm},\tilde \psi^{\pm})$ 
of the fermionic $SU(2)$ that we discussed before, or equivalently to the compact boson of the cigar which 
is asymptotically at the fermionic radius, for which the twisted sectors of the $\zi_p$ orbifold translate 
into fractional windings $\gamma /p$. The relative sign of the orbifold action on the left- and right- moving sector is such 
that the $\cn = 2$ Liouville operator is not projected out. From the geometrical point of view it means that the orbifold 
has to act in the $(x^8,x^9)$ plane and not in the $(x^6,x^7)$ one.

Now we can add a D4-brane stretched between the two NS5-branes, 
localized in the plane $(4,5)$. Without the orbifold, the one-point function 
for the D4-brane in type \textsc{iia} reads~\cite{Israel:2005fn,Eguchi:2003ik}, 
(in the light-cone gauge)
\begin{eqnarray}
\langle \ V_{\tilde \jmath,{\bf p}}^{(s_i)\
 (\bar s_i)}\
 \rangle_{\hat{s}_i,{\bf \hat{y}}}
=   |z-\bar z |^{-\Delta-\bar \Delta}\ 2^{-\frac{1}{2}-\tilde \jmath} \ 
\delta_{s_1,-\bar{s_1}} \delta_{s_2, -\bar{s_2}} \delta_{s_3,-\bar{s_3}}
\delta_{s_4,-\bar{s_4}} \, \delta^{(2)} ({\bf p} ) \ \times \hphantom{aaaaaaaaa}
\nonumber\\ \times \quad   e^{i (p_4 \hat y^4 +p_5 \hat{y}^5)}
e^{i\frac{\pi}{2} \sum_{i} s_i \hat{s}_i }\
\frac{\Gamma \left( \tilde \jmath + \frac{s_4-s_3}{2}
\right) \Gamma \left( \tilde \jmath + \frac{\bar s_4 -\bar  s_3}{2} \right)}{\Gamma (2\tilde \jmath-1)
\Gamma ( \frac{1}{2}+\tilde \jmath  )}\, ,\nonumber\\
\label{oneptW}
\end{eqnarray}
for a closed string vertex operator with $\slr$ spin $\tilde \jmath$ (i.e. of the form $V \sim e^{-Q\tilde \jmath \phi}$) fermionic R-charges $(s_i,\bar s_i)$ for $i=1,\ldots,3$ (the 
last ones being the fermions of the cigar). 
The charges $(s_4,\bar s_4)$ correspond to the left and right chiral momenta of the 
compact boson in the cigar $\slc$. 
This one-point function gives the following open string annulus amplitude for open strings with both ends 
on the same D-brane:
\begin{equation}
Z_\textsc{open} (\tau) = \frac{1}{-i\tau 8\pi^2 \alpha' \eta^4} \ 
 \frac{1}{2} \sum_{a,b=0}^1  \sum_{\{ \upsilon_\ell \} \in (\zi_2)^4}(-)^{a+b(1+\sum_\ell \upsilon_\ell)} 
\ \frac{\Theta_{a+2\upsilon_1,2}\Theta_{a+2\upsilon_2,2}}{\eta^2}\  
Ch_\mathbb{I}^{(a+2\upsilon_3)} \left( \upsilon_4 ;\tau \right)\, .
\end{equation}
The massless spectrum corresponds to the dimensional reduction of a six-dimensional \mbox{$\cn =1$} gauge multiplet 
to four dimensions.  For $n=pp'$ coincident D4-branes, one obtains 
an $\cn = 2$ $U(n)$ gauge multiplet.
It contains, from the \textsc{ns} sector, a four-dimensional gauge field, given by 
\begin{equation}
\lambda_\textsc{g}\ A_\mu \psi^{\mu}_{-1/2}\ |0 \rangle_\textsc{ns}\ \otimes \ | p^\mu \rangle \ \otimes \ 
|r=0 \rangle_\textsc{sl(2,r)/u(1)}  \ ,
\end{equation}
and a complex scalar 
\begin{equation}
 \lambda_\textsc{c}\ (\xi^{4} \pm i\xi^5)_{-1/2}\ |0 \rangle_\textsc{ns}\ \otimes \ | p^\mu \rangle  \ 
 \otimes \ |r=0 \rangle_\textsc{sl(2,r)/u(1)}   \ .
 \label{chirmult}
\end{equation}
The $pp' \times pp'$ matrices $\lambda_\textsc{g}$ and $\lambda_\textsc{c}$ are associated with 
the Chan-Paton factors. Before the orbifold they are both arbitrary Hermitian matrices, 
associated to the adjoint representation of $U(pp')$. There are no other massless degrees of freedom since 
all the excitations along the $\slc$ factor are massive. 
Now we perform the $\zi_p$ orbifold of the open string theory. In particular we have to find 
the action of the orbifold group onto the matrices associated to the Chan-Paton factors. 
The solution to this problem is quite similar to the construction of D-branes transverse to an ALE space~\cite{Douglas:1996sw,Johnson:1996py}, in the present case to an $A_{p-1}$ singularity, 
since the action of the $\zi_p$ orbifold is Abelian. The action of the 
orbifold on the $\slc$ open string operators in the \textsc{ns} sector ($a=0$),  
will be only on the massive states of the open string theory since the vacuum $r=0$ of $\slc$ is 
invariant. One obtains then,from the action on the $x^{4,5}$ directions and on 
the Chan-Paton factors, an $\cn = 1$ quiver $U(p')^{p}$, for 
$pp'$ D4-branes on the covering space of the orbifold. The excitations along $x^{4,5}$ turn into 
bifundamental chiral multiplets $Q_{\ell \to \ell+1}$ connecting the nodes of the affine $\hat{A}_{p-1}$ Dynkin diagram. 
This quiver is a {\it chiral} one since there is only one chiral multiplet associated to each link. Then 
as explained above the $U(1)$ are all anomalous and we end up with an $SU(p')^p$ gauge group. 

From the construction of the gauge theory using boundary worldsheet \textsc{cft} outlined above we obtain 
only the classical properties of the field theory. The perturbative corrections can be found 
by computing the bending of the NS5-branes --~corresponding to the one-loop running 
of the coupling constant~-- as a closed string tadpole for the 
string states holographically dual to the operators $\mathrm{Tr} (X^6+iX^7)^r$ 
on the fivebranes worldvolume~\cite{Israel:2005fn}. It would be also interesting to compute the $U(1)$ anomalies 
from the worldsheet boundary \textsc{cft} approach; they should correspond to tadpoles in the \textsc{r-r} 
twisted sectors of the $\zi_p$ orbifold. To describe the full quantum theory in the strong coupling 
regime one needs to go to the four-dimensional non-critical strings regime, 
for which this configuration corresponds  to a single NS5-brane wrapping a singular Riemann surface 
embedded in $\mathbb{C}\times \mathbb{C}/\zi_p$. 

\boldmath
\section{Double scaling limit of $\cn = 1$ quivers}
\unboldmath
\label{n1duals}
We have demonstrated above that an asymmetric orbifold of $\cn =2$ non-critical strings 
leads to an $\cn = 1$ string theory, holographically dual to a non-gravitational 
theory flowing at low energies to an $\cn = 1$ affine $\hat{A}_{p-1}$ quiver gauge theory, in the Coulomb phase. 
\FIGURE{
\raisebox{2.5cm}{
\begin{picture}(50,60)
\thicklines
\put(7,5){\circle{14}}
\put(8,-10){\makebox(0,0){\tiny $SU(p p')$}}
\end{picture}
\hskip1cm
$\stackrel{\zi_{p}}{\longrightarrow}$}
\hskip1cm
\raisebox{2.5cm}{{\scriptsize $p-1$ nodes} 
$\left\{ \phantom{\begin{array}{c}
a \\ a\\ a\\ a\\ a\\ a\\ a\\ a\\ a\\ a \\ a
\end{array}}
\right.$
}
\begin{picture}(50,150)
\thicklines
\put(7,5){\circle{14}}
\put(25,-10){\makebox(0,0){\tiny $SU(p ')$}}
\put(7,12){\vector(0,1){28}}
\put(7,47){\circle{14}}
\put(25,32){\makebox(0,0){\tiny $SU(p ')$}}
\put(7,54){\vector(0,1){28}}
\put(7,89){\circle{14}}
\put(25,74){\makebox(0,0){\tiny $SU(p ')$}}
\put(7,96){\dashbox{3}(0,45)}
\put(7,148){\circle{14}}
\put(25,133){\makebox(0,0){\tiny $SU(p ')$}}
\put(80,75){\circle{14}}
\put(80,53){\makebox(0,0){\tiny $SU(p ')$}}
\put(74,70){\vector(-1,-1){61}}
\put(12,142){\vector(1,-1){62}} 
\end{picture}\\
\caption{  Quiver diagram obtained by a $\zi_{p}$ orbifold of the $SU(p p' )$ theory.}
\label{quivdiagsimp}}
To each node of the quiver is associated an $\cn = 1$ 
vector multiplet and to each link a bifundamental $\cn = 1$ chiral multiplet $Q_{\ell \to \ell+1}$ transforming in 
$(\bar \Box, \Box)$, see fig.~\ref{quivdiagsimp}.\footnote{  
Note that this gauge theory can also be obtained by deconstruction of five-dimensional 
\textsc{sym}~\cite{Csaki:2001zx,Iqbal:2002ep,DiNapoli:2004rm}.} Since there is only one oriented link between 
adjacent nodes, i.e. chiral multiplets transforming in $(\bar \Box, \Box)$ and not 
in $(\Box, \bar \Box)$, this quiver gauge theory is chiral. 
It has no superpotential, because there is no superpotential inherited from the projection 
of the $\cn = 2$ pure gauge theory --~in opposition with what happens for $\cn =1$ gauge theories coming from orbifolds  
of $\cn = 4$ theories~\cite{Kachru:1998ys}. The absence 
of superpotential could be checked from the 
worldsheet boundary \textsc{cft} construction of the previous section, since each term in the superpotential can 
be computed as open string $n$-point functions of the appropriate boundary fields, as was done in~\cite{Brunner:2000wx} 
for Gepner models. The action of the orbifold on the open string field theory, in particular on the chiral multiplets~(\ref{chirmult}), 
won't generate new nonzero correlators.

From our construction it is quite clear that the linear dilaton limit of the non-critical string 
corresponds to a non-trivial $\cn = 1$ superconformal field theory on the gauge theory side. 
This superconformal gauge theory is an Argyres-Douglas point 
in the Coulomb phase of the affine $\hat{A}_{p-1}$ quiver, where the gauge symmetry is generically 
$U(1)^{p'-1}$. As for the $\cn = 2$ models we will also study the double scaling limit 
of the non-critical string dual, corresponding to the neighborhood of the 
superconformal fixed point. 

\subsection{Chiral ring of the quiver}
In order to understand the matching between the chiral ring of the quiver and the string theory massless 
operators that are chiral in spacetime, we start by describing the moduli space of the 
gauge theory and the associated \textsc{sw} curve, following~\cite{Csaki:1997zg,Hailu:2002bg}. 
A basis of independent gauge-invariant generators of the chiral ring (omitting for the moment 
the operators involving the gaugino superfield) is provided by
\begin{equation}
\begin{array}{rclr}
\mathfrak{L}_r &=& \mathrm{Tr} (Q_{1\to 2} Q_{2\to 3} \cdots Q_{p\to 1} )^r  \ , &\qquad r=1,\ldots,p'-1  \\
\mathfrak{B}_{\ell \to \ell +1} &=& \det Q_{\ell \to \ell +1} \ ,  & \ell \sim \ell + p
\end{array}
\label{chiring}
\end{equation}
The first type of operators are made from the "loop" operator, i.e. the quiver-ordered product of all 
the link fields. It can be used to define a composite adjoint operator
\begin{equation}
\Xi = Q_{1\to 2} Q_{2\to 3} \cdots Q_{p\to 1} -\frac{1}{p'} \, \mathfrak{L}_1 \ \mathbb{I}_{p' \times p'} \, .
\end{equation} 
The second type of operators are "baryons" and are the generators of a larger class of 
determinants of products of link fields $\mathfrak{B}_{\ell \to \ell + p} =  \det (Q_{\ell \to \ell+1} 
\cdots Q_{\ell+p-1 \to \ell+p})$. The classical constraint between those and the "elementary" baryons operators is 
modified quantum mechanically~\cite{Seiberg:1994bz}. In particular using such a relation we can 
express \mbox{$\mathfrak{L}_{p '} = \mathrm{Tr}\, (Q_{1\to 2} \cdots Q_{p\to 1} )^{p'}$} in terms 
of the other gauge-invariant operators, as follows~\cite{Rodriguez:1997ng,Chang:2002qt}
\begin{subequations}
\begin{align}
\mathfrak{B}_{1\to 1} &= \mathfrak{B}_{1 \to 2} \times \cdots \times  \mathfrak{B}_{p \to 1} + \left( 
\mathfrak{B}_{r \to r + 1} \mathfrak{B}_{r+1 \to r + 2} \to \Lambda_{r+1}^{2p'}
\right) \label{detrelA} \\
&= \frac{(-1)^{p'+1}}{p'}\ \mathfrak{L}_{p'} +\sum_{r=2}^{p'} \sum_{\{n_i\} =1}^{p' -1} \delta_{\sum n_i,p'}\frac{(-)^{r+p'}}{r!} 
\frac{\mathfrak{L}_{n_1}}{n_1} \cdots \frac{\mathfrak{L}_{n_r}}{n_r}
\label{detrelB}
\end{align} 
\label{detrel}
\end{subequations}
where on the right-hand side of~(\ref{detrelA}) we mean that we must add all the possible terms obtained by 
replacing pairs of adjacent baryonic operators $(\mathfrak{B}_{r \to r + 1} \mathfrak{B}_{r+1 \to r + 2})$
by $\Lambda^{2p'}_{r+1}$, corresponding to the dynamically generated scale 
for the $(r+1)$-th $SU(p')$ gauge group on the quiver diagram. 

The properties of the Coulomb phase for $\cn = 1$ gauge theories are given, as for the $\cn = 2$ models in terms 
of an hyperelliptic curve~\cite{Intriligator:1994sm}, which allows to compute the gauge coupling $\tau$ 
as a function of the coordinates on the moduli space. For the $SU(p')$ quivers associated to the $\hat{A}_{p-1}$ 
diagram that we consider, the curve is given by 
\begin{equation}
y^2 = \frac{1}{4} \langle \det (x-\Xi ) \rangle^2 - \Lambda_{1}^{2p'} \cdots \Lambda_{p}^{2p'}
\end{equation}
Using the relation~(\ref{detrel}) the parameter $s_{p'} (\Xi )$ of this curve can be related to the vacuum expectation values 
of the chiral ring operators~(\ref{chiring}).  In our particular 
gauge theory derived from a string theory construction, the vacuum expectations values of all the baryonic operators are related by D-terms constraints, coming from the anomalous $U(1)$ gauge symmetries discussed in the previous section. Because of the 
$\zi_{p}$ symmetry of the theory we also consider all the scales $\Lambda_\ell$ to be the same. 
This curve have the same type of singularities 
as the $\cn = 2$ $SU(p')$ curve that we discussed in the previous sections. In terms of the parameters~(\ref{chiring}) these singularities are not points but rather hypersurfaces of the moduli space. They correspond to the same number 
of dyons becoming massless, although their mass cannot be determined a priori since they are non-holomorphic 
quantities. In particular, for $p' >2$ the moduli space contains singularities of the Argyres-Douglas type 
where mutually non-local dyons become massless, leading to a strongly interacting $\cn = 1$ superconformal 
field theory in the infrared. As for the $\cn = 2$ models, the string dual describe these singularities and 
--~when the strong coupling singularity of the linear dilaton is resolved~-- small perturbations 
of those.

Now that we have a clear picture of the moduli space of the Coulomb branch, we can find the mapping between 
the gauge-invariant chiral operators~(\ref{chiring}) of the quiver and string massless states that are chiral 
in spacetime. First we can identify the invariant polynomials of the composite adjoint $\Xi$, defined in 
eq.~(\ref{polchar}),  with string \textsc{ns-ns} chiral primaries from the untwisted sectors:
\begin{equation}
\mathcal{V}_{\frac{pN}{2}}^\textsc{u} = 
e^{-\varphi-\tilde \varphi} e^{-Q \tilde \jmath \rho} 
e^{i \sqrt{\frac{p}{p' (pp'+2)}} N (X_L - X_R)} 
V_{\frac{pN}{2}-1, \,\frac{pN}{2}, \, \frac{pN}{2}}^{(2,2)}  \ \longleftrightarrow \ 
s_{N} (\Xi) \ , \quad  N=0, \ldots , p'-1 \, .
\end{equation} 
Only the operators with $N>p' /2 +1/p$ will correspond to 
string theory observables, and from the gauge theory point of view they 
satisfy the unitarity bound at the superconformal fixed point.  
It gives the following holographic prediction for the scaling dimensions of
such chiral operators
\begin{equation}
\Delta [ s_{N} (\Xi) ] \ =  \ \frac{4pN}{pp'+2}
\end{equation}
with the same issue of mixing as discussed in the $\cn = 2$ case, see eq.~(\ref{mixop}). 
This spectrum of dimensions is different from the $\cn = 2$ theory corresponding to the diagonal
$SU(p')$ of the quiver which is not Higgsed (but broken to $U(1)^{p'-1}$ on the Coulomb branch). 
This seems surprising since the two theories share the same type of \textsc{sw} curve. In 
the $\cn = 2$ theories, the overall normalization of the R-charges for the perturbations around the singularity 
can be found by requiring that the K\"ahler potential has 
dimension two, which implies that the Seiberg-Witten one-form has dimension one~\cite{Argyres:1995xn}. 
However, for the $\cn = 1$ theories, the K\"ahler potential is not given by the curve and such a 
constraint does not apply. 

As we saw the chiral ring contains also "baryonic" operators made of determinants 
of link fields. They are identified with 
\textsc{ns-ns} worldsheet chiral primaries in the twisted sector of  the string theory orbifold 
as follows 
\begin{equation}
\mathcal{V}_{\frac{p'\gamma}{2}}^{\textsc{t}} = 
e^{-\varphi-\tilde \varphi} e^{-Q \tilde \jmath \rho} 
e^{i  \sqrt{\frac{p'}{p (pp'+2)}}\, \gamma\, (X_L - X_R)} 
V_{\frac{p'\gamma}{2} -1 ,\,  -\frac{p'\gamma}{2}, \, \frac{p'\gamma}{2}}^{(2,2)} 
\ \longleftrightarrow  \  
\mathfrak{B}_{1 \to \gamma } \, .
\end{equation} 
One may be worried by the fact that the left-hand side of the dictionary seems to be dependent of the 
position on the quiver diagram, while the string vertex operator one the right-hand 
side doesn't carry any associated label. What happens is that the D-terms 
coming from the anomalous U(1)'s will give $p-1$ constraints relating the $p$ different baryonic 
operators of~(\ref{chiring}), see~\cite{Lykken:1997gy}. Another issue is whether 
the gauge-theory chiral operator is a composite baryon $\mathfrak{B}_{1\to \gamma }$ 
or a product of elementary baryons $\mathfrak{B}_{\ell \to \ell+1 }$. These two options 
differ from terms involving lower powers of the elementary baryons, using relations 
like~(\ref{detrelA}) between the composite and elementary baryons. We should 
also not forget that, as for the untwisted sector, there is some operator mixing 
at the level of the holographic dictionary. Anyway we get a prediction from the string theory side of 
the scaling dimensions of the baryonic chiral operators at the superconformal fixed point 
\begin{equation}
\Delta (\mathfrak{B}_{1 \to \gamma} ) \ = \ \frac{4p'
  \gamma }{pp'+2} \, .
\end{equation} 
The operator with $\gamma = p$, i.e. 
\begin{equation}
\mathfrak{B}_{1 \to 1} = \det Q_{1\to 2} Q_{2\to 3} \cdots Q_{p\to 1} \, ,
\end{equation} 
is given on the string theory side by the $\cn=2$ Liouville potential. It can viewed either as belonging 
to the twisted or to the untwisted sector from the string theory point of view, which is consistent 
via the holographic dictionary to the chiral ring that we discussed above. 
This analysis carries over straightforwardly to the operators in the \textsc{r-r} sector. 
The \textsc{r-r} ground states of the untwisted sector correspond to the chiral operators 
(in $\cn = 1$ superfield notation) $\mathrm{Tr} \, W^\alpha_1 (Q_{1\to 2} \cdots Q_{p \to 1})^{r}$.

The Argyres-Douglas point for which the curve is the most singular --~or in other words for which 
the biggest number of dyons become massless~-- is obtained, taking for example an 
$SU(3)^p$ quiver, whose curve depends on the polynomial $P_3 (x) = 
x^3-\tilde u x - \tilde v$, by the hypersurface
\begin{eqnarray}
\tilde u &=& \frac{1}{2} \mathrm{Tr}\, \langle \Xi^2 \rangle = 
\frac{1}{2} \langle \mathfrak{L}_2 \rangle- \frac{1}{6}  \langle \mathfrak{L}_1^{\, 3} \rangle = 0 \, , \nonumber \\
\tilde v &=& \frac{1}{3} \mathrm{Tr}\,\langle \Xi^3\rangle = \langle \mathfrak{B}_{1\to 1}\rangle + \frac{1}{6}
 \langle \mathfrak{L}_1  \mathfrak{L}_2 \rangle - \frac{5}{54}  \langle \mathfrak{L}_1^{\, 3} \rangle = \pm 2 \Lambda^{pp'} \, .
\end{eqnarray}
According to our dictionary, the double scaling limit of the string dual that 
we consider corresponds to the submanifold of the moduli space for which the only operator with 
non-zero vacuum expectation value away from the \textsc{ad} point is 
$\mathfrak{B}_{1 \to 1}$, giving rise to the $\cn = 2$ Liouville perturbation in the string theory.

\subsection{Dyon spectrum and stable D-branes}
The various singularities of the \textsc{sw} curve of the $\cn = 1$ gauge theory correspond 
to dyons becoming massless as for the $\cn =2$ theories. These dyons are of course not \textsc{bps} anymore, and 
their domain of stability is unknown; nevertheless in the double scaling limit we describe 
a small neighborhood of the singularity where we can assume that they are stable. In the string theory 
dual they are naturally related to the localized D-branes in the $\cn = 1$ little string 
theory background.

These D-branes can be constructed in close parallel with the $\cn=2$ models, because the orbifold 
acts freely.  Then the one-point function on the disc for localized D-branes 
is the same as~(\ref{onept}), up to an overall normalization, however now the closed string operators that can appear 
are restricted by the orbifold condition. In other words, the boundary state is constructed 
from a restricted set of Ishibashi states. The twisted sectors, containing the 
baryonic operators in the massless \textsc{ns-ns} spectrum, don't couple to the D-brane, so 
we have simply to keep the closed string operators with $m = 0 \mod p$. 
Then we can compute the open string partition function
\begin{eqnarray}
Z_{\textsc{open}} (\tau ) =  
\frac{q^{\left( \frac{\hat{\mathbf{y}}' -\hat{\mathbf{y}}}{2\pi} \right)^2}}{
\eta^2} \sum_{\{ \upsilon_i\} \in (\zi_2)^3} 
\frac{1}{2} \sum_{a,b \in \zi_2} (-)^{a+b(1+\upsilon_1+\upsilon_2+\upsilon_3+m)} 
\frac{\Theta_{a+2\upsilon_1+\hat s_1 - \hat s_1 ',2}}{\eta} 
\ \times 
\nonumber \\ \times \ 
\sum_{2j = 0}^{n-2} \sum_{m \in \zi_{2n}} \ N^{j}_{\hat \jmath \hat \jmath '}  \  \sum_{\omega \in \zi_p}  
C^{j\, (a+2\upsilon_3+\hat s_3 - 
\hat s_3 ')}_{2m+a+\hat m - \hat m'+2p' \omega} \ 
Ch_{\mathbb{I}}^{(a+2\upsilon_2+\hat s_2 -\hat s_2 ')} (m)  \, .
\nonumber \\
\label{dyonbrane}
\end{eqnarray}
We see that spacetime supersymmetry is broken by the sectors with $\omega \neq 0$. Also because of them 
the $\zi_{pp'}$-valued label $\hat m$ of the brane is now restricted to $\hat m \in \zi_{2p'}$ 
(otherwise it gives equivalent open string sectors). 

Let's consider open strings with both endpoints on the same D-brane. 
The first type of strings  are given by $\upsilon_1=1$ and $\upsilon_2=\upsilon_3=0$. 
Then the sector $\omega = 0$ contains massless states in spacetime, corresponding to a 
four-dimensional gauge field reduced to 0+1 dimension, as in the $\cn = 2$ case~(\ref{gaugefopen}). 
In the Ramond sector we have similarly a four-dimensional massless spinor; 
altogether they form an $\cn =1$ gauge multiplet reduced to 0+1 dimensions. 
The sectors $\omega \neq 0$ for $\upsilon_1 = 1$ are all massive. 
However we have potentially new types of light states 
with a fermionic excitation along $SU(2)/U(1)$, i.e. $\upsilon_3=1$ and $\upsilon_1=\upsilon_2=\upsilon_4=0$. 
In all the sectors with $\omega \neq 0$, one can construct a worldsheet 
chiral primary of the worldsheet $\cn = 2$ \textsc{sca}, 
provided the open string spectrum contains a representation of spin $j$ which satisfies $j+1=p'\omega$. 
Then for $m=0$ we find a open string physical state with mass 
\begin{equation}
m_\textsc{Tach}^2 = -\frac{\omega}{\alpha ' p} \, ,
\end{equation}
thus it is an open string tachyon and this D-brane will decay. To avoid these tachyons in the 
various open string sectors with both endpoints on the same D-brane, 
one needs to consider only the subset  of 
D-branes with $SU(2)$ labels 
\begin{equation}
\hat \jmath \leqslant  \frac{p'}{2}-1\, ,
\label{trunc}
\end{equation}
such that from the fusion rules $N^{j}_{\hat \jmath \hat \jmath '}$ 
we never have representations with $j+1 \geqslant p'$. One can check that for 
values of $j$ lower than $p'$ there are no tachyons either for the non-chiral 
states of $SU(2)/U(1)$. We conclude that the spectrum of open strings with both ends 
on any D-brane of this truncated set is non-tachyonic, thus they are stable in string theory.  
\FIGURE{ \hskip2cm \epsfig{figure=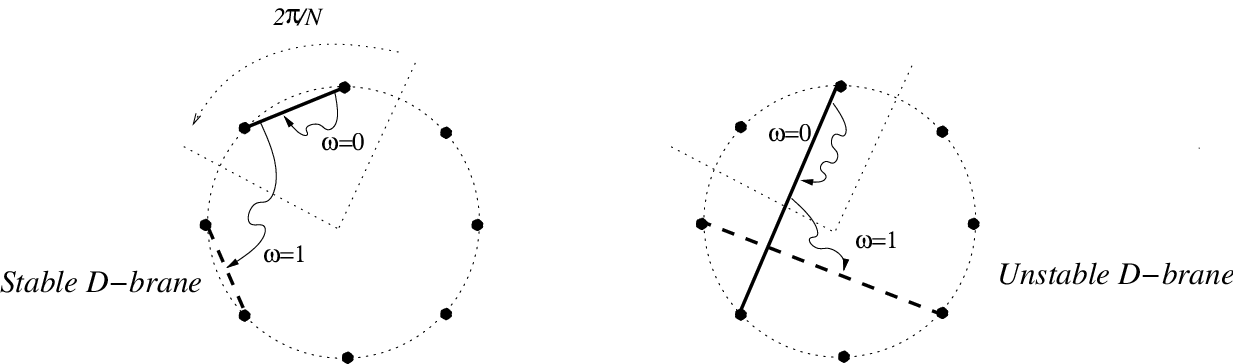, width=120mm} \caption{
Stable D-brane (left) and unstable one (right) in $\cn =1$ non-critical strings.}
\label{branesorb}}
These results can be understood geometrically, see fig.~\ref{branesorb}. 
At the level of the localized D-branes construction, the action of the orbifold 
on the $SU(2)/U(1)$ coset can be thought as a $2\pi/p$ rotation around the 
center of the coset target space metric, which is conformal to a disc; the 
fundamental domain of the orbifold contains $p'$ of the "special points" on 
the boundary of the disc on which the D1-branes of the coset can end.  
The D-branes with $\hat \jmath \leqslant p'/2-1$ fit into the fundamental 
domain of the orbifold, so they won't intersect with their images and the 
open string sectors between different images --~that break supersymmetry~-- 
will be very massive. On the contrary for D-branes that don't fit in the fundamental domain 
there is an unstable mode associated with the recombination of the intersections with their 
images.

This is compatible with the gauge theory expectations. 
Indeed since the Seiberg-Witten curve describing the Coulomb branch of the $\hat{A}_{p-1}$, $SU(p')$ quiver 
is quite similar to the curve of the $\cn = 2$ pure $SU(p')$ theory,  the theory must contain the 
same number of light, stable dyons near the Argyres-Douglas point. The truncation of the allowed 
D-branes in the $\cn = 1$ model~(\ref{trunc}), together with the 
restriction $\hat m \in \zi_{2p'}$ already discussed,  give precisely the same number of D-branes 
as the $\cn = 2$ non-critical string dual to the $\cn = 2$ $SU(p')$ theory. 
We have seen also that the massless degrees of freedom on the D-branes correspond to 
an $\cn = 1$ gauge multiplet; by considering sectors of open strings between different 
stable D-branes we can compute the open string Witten index that will be the same as for 
D-branes in the dual of the $\cn = 2$ $SU(p')$ theory. This is not surprising since the charges 
of the dyons becoming massless can be computed from the monodromies of the Seiberg-Witten 
curve, which is the same for both gauge theories. 

From the one-point function with a graviton vertex operator we can again compute 
the masses of these dyons.  The value of the cigar perturbation
is related to the masses of these stable localized D-branes which are the lightest 
non-perturbative states of the string theory, as in the $\cn = 2$ models. 
However, from the gauge theory side, these masses are not given anymore 
by the Seiberg-Witten curve of the theory since the dyons are not \textsc{bps}.
From the string theory dual we can see that the masses have the following scaling relation with 
respect to the moduli space coordinate $\tilde v$ of the $\cn = 1$ quiver corresponding to the $\cn = 2$ Liouville
perturbation:
\begin{equation}
m_\textsc{dyon}^2 \sim (\delta \tilde v)^{\frac{pp'+2}{pp'}} \, ,
\end{equation}
which depends only on the product $pp'$. 
Although the physics of this $\cn = 1$ quiver in the Coulomb phase is quite similar to 
an $\cn = 2$ $SU(p')$ theory, we see that the scaling relation between the dyon mass and 
the deformation of the singularity is different. 
We can compute from the string theory dual the ratio of masses for dyons of labels $\hat \jmath $ and $\hat \jmath '$:
\begin{equation}
\frac{m_{\hat \jmath}}{m_{\hat \jmath '}}
 = \frac{\sin \frac{\pi (2\hat \jmath+1)}{pp'}}{
\sin \frac{\pi (2\hat \jmath '+1)}{pp'}} \quad \text{for} \qquad 0 \leqslant \hat \jmath, \, \hat \jmath ' 
\leqslant \frac{p'}{2}-1 \, . 
\end{equation}
This is a prediction of the string theory dual that cannot be matched to a similar quantity 
in the gauge theory, since the tools for computing it are lacking.\footnote{It is natural in the related 
context of \textsc{AdS}/\textsc{cft}, where we consider orbifolds of $\cn =4$ \textsc{sym}, 
that the projected theory shares many properties with the original one.}  
However, as again these ratios are independent of the double scaling parameter we expect that
they are exact results in the gauge theory.

\section{Conclusions}
In this work we studied some non-critical string duals of four-dimensional 
supersymmetric gauge theories that are exactly solvable worldsheet conformal field theories. 
As for all the linear dilaton backgrounds the dual, non-gravitational theory 
is not a field theory, rather a little string theory, whose low-energy physics is nevertheless 
well described by a certain gauge theory. It is not possible 
to have a neat separation between the energy scales of the gauge theory and the energy scale 
of the little string theory --~above which non-field theoretic excitations mix with the 
gauge field theory~-- while keeping the string coupling constant small. Nevertheless 
worldsheet quantities having to do with the localized states (loosely speaking, 
excitations at the tip of the cigar) should correspond to field theory physics, and indeed 
are in good agreement with field theory expectations. 

The first examples, that were already studied~\cite{Giveon:1999zm,Pelc:2000hx,Eguchi:2004ik}, 
of such models are $\cn = 2$ non-critical string duals of $SU(n)$ gauge theories near an Argyres-Douglas 
point where the field theory is superconformal. The string/gauge duality then allowed 
to compute the anomalous dimensions of chiral operators at the superconformal 
fixed point using the partition function 
of the string theory that we obtained. 
We constructed explicitly the \textsc{bps}, localized D-branes that are dual to 
the \textsc{bps} dyons of the gauge theory at the superconformal fixed point, 
and show that their masses can be computed in terms of the coordinates of the moduli space 
of the Coulomb branch using the $\slc$~--~$\cn =2$ Liouville duality which 
is a worldsheet instanton effect. In other words 
this shows that worldsheet non-perturbative effects give the non-trivial Seiberg-Witten 
solution of the gauge theory. 

We have shown that this duality between non-critical strings and gauge theories can be 
extended to theories with only $\cn = 1$ supersymmetry. More precisely, a chiral orbifold 
of the $\cn = 2$ non-critical strings  considered above gives $\cn = 1$ non-critical 
strings dual to $\cn =1$ chiral quivers 
in the Coulomb phase, whose moduli space contains also Argyres-Douglas superconformal 
fixed points. These are to our knowledge the first examples of string duals of 
$\cn = 1$ gauge theories that can be solved exactly, at the level of string perturbation 
theory. The spectrum of scaling dimensions for the chiral ring operators at the superconformal 
fixed point can be also computed in those examples. It is very interesting also that 
the dyons that become massless at the superconformal fixed point are in one-to-one 
correspondence with non-supersymmetric D-branes which are nevertheless stable. Indeed 
their massless spectrum is supersymmetric in spacetime. The predictions of the 
string duals for the masses of those dyons are quite interesting since similar quantities 
cannot be computed in the field theory. 

There are obvious generalizations of these models, if we consider the
more generic class of four-dimensional non-critical strings~(\ref{back4lst}) with 
two $\cn = 2$ minimal models. These are dual to $\cn = 2$ quivers, and further 
orbifoldization leads to more complicated $\cn = 1$ quivers with interesting 
structure. We will study them in a forthcoming publication. Another interesting 
feature of these models is that, since they are intrinsically ten-dimensional 
superstring theories, they have a clear low-energy supergravity limit. 
From all these examples we see an interesting relation between critical phenomena 
in two dimensions, i.e. two-dimensional \textsc{cft}s, and critical points  
of four-dimensional gauge theories. It would be interesting to understand how 
far this correspondence can be extended.

Another development of this work can be to construct, from both types 
of non-critical strings, four-dimensional $\cn= 1$ gauge theories by adding 
D-branes extended in the four spacetime dimensions, that are T-dual (along 
the flat space directions) to the 
"dyonic" D-branes discussed above, generalizing the 
construction with D-branes in the conifold~\cite{Fotopoulos:2005cn,Ashok:2005py}.  
The D-branes of the $\cn = 1$ non-critical 
strings give in particular the possibility of obtaining 
supersymmetric gauge theories from non-supersymmetric D-branes.  

The most interesting and challenging issue is to understand how to 
add tree level superpotentials to the gauge theories considered in this 
paper. From the string theory point of view it should correspond to 
adding non-normalizable marginal deformations to the worldsheet action, which 
raises many puzzles. If we trust the field theory picture at the 
level of the string theory --~even though the non-critical string 
is not strictly speaking dual to the gauge theory~-- we expect 
to find a large variety of interesting phenomena. Indeed, from the 
gauge theory point of view, adding a tree level superpotential to an 
$\cn = 2$ or $\cn =1$ theory at a critical point in the Coulomb phase 
can lead either to a lifting of this vacua, confinement through 
condensation of monopoles or an $\cn = 1$ superconformal field theory. 
These three possibilities would correspond respectively in the string 
theory to an instability (leading to a time-dependent solution 
pushing the theory away from the critical point), condensation of 
D-branes generating Ramond-Ramond flux and, if we start 
with the $\cn = 2$ non-critical strings, breaking of $\cn = 2$ to 
$\cn = 1$ by a marginal non-normalizable deformation.

\section*{Acknowledgements}
I would like to thank N.~Dorey, A.~Hanany, J.~Sonneschein, and especially S.~Elitzur and A.~Giveon 
for very helpful discussions. This work is supported by a European Union 
Marie Curie Intra-European Fellowship under the contract MEIF-CT-2005-024072, a European Union 
Marie Curie Research Training Network under the contract MRTN-CT-2004-512194, the American-Israel Bi-National Science 
Foundation and the Israel Science Foundation.

%%%%%%%%%%%%%%%%%%%%%%%%%%%%%%%%%%%%%%%%%%%%%%%%%%%%%%%%%%%%%%%%%%%%%%%%%%%%%%%%%%%%%%%%%%%%%%%%%%%%%%%%
%%%%%%%%%%%%%%%%%%%%%%%%%%%%%%%%%%%%%%%%%%%%%%%%%%%%%%%%%%%%%%%%%%%%%%%%%%%%%%%%%%%%%%%%%%%%%%%%%%%%%%%%
%%%%%%%%%%%%%%%%APPENDICES%%%%%%%%%%%%%%%%%%%%%%%%%%%%%%%%%%%%%%%%%%%%%%%%%%%%%%%%%%%%%%%%%%%%%%%%%%%%%%
%%%%%%%%%%%%%%%%%%%%%%%%%%%%%%%%%%%%%%%%%%%%%%%%%%%%%%%%%%%%%%%%%%%%%%%%%%%%%%%%%%%%%%%%%%%%%%%%%%%%%%%%

\appendix
\boldmath
\section{Representations and characters 
of worldsheet $\cn = 2$ superconformal algebra}
\unboldmath

We gather in this appendix some conventions and modular properties
of characters that we use abundantly in the bulk of the paper,
and in particular in the transformations of the annulus amplitude from
open to closed string channels.
\subsection*{Free fermions}
We define the theta-functions as:
$$\vartheta \oao{a}{b} (\tau,\nu )
= \sum_{n \in \zi} q^{\frac{1}{2} (n+\frac{a}{2})^2}
e^{2 i \pi (n+\frac{a}{2})(\nu+\frac{b}{2})},$$
where $q=e^{ 2 \pi i \tau}$. 
It will be convenient to split the states inside the R and NS sectors
according to their fermion number:
\begin{equation}
\begin{array}{ccllll}
 \frac{1}{2\eta} \left\{ \vartheta \oao{0}{0} - \vartheta \oao{0}{1} \right\} & = &
\frac{\Theta_{0,2}}{\eta}
\\
 \frac{1}{2\eta} \left\{ \vartheta \oao{0}{0} + \vartheta \oao{0}{1} \right\} & = &
\frac{\Theta_{2,2}}{\eta}
\\
 \frac{1}{2\eta} \left\{ \vartheta \oao{1}{0} - i\vartheta \oao{1}{1} \right\} & = &
\frac{\Theta_{1,2}}{\eta}
\\
\frac{1}{2\eta} \left\{ \vartheta \oao{1}{0} + i\vartheta \oao{1}{1} \right\} & = &
\frac{\Theta_{3,2}}{\eta}
\\
\end{array}
\end{equation}
in terms of the theta functions of $\hat{\mathfrak{su}} (2)$
at level $2$:
$$\Theta_{m,k} (\tau,\nu) = \sum_{n \in \zi}
q^{k\left(n+\frac{m}{2k}\right)^2}
e^{2i\pi \nu k \left(n+\frac{m}{2k}\right)}.$$ 
The modular transformation property of the fermionic characters is then:
\begin{equation}
\frac{\Theta_{s,2} (-1/\tau , \nu/\tau )}{\eta (-1/\tau  )}  = \frac{1}{2} e^{i\pi \nu^2 /\tau}
\sum_{s' \in \zi_4}
e^{-i\pi ss'/2} \frac{\Theta_{s',2} (\tau , \nu)}{\eta (\tau )}  .
\end{equation}
We will often work in terms of these characters for fermions in NS or
R sector, projected onto even or odd fermion number states.

\boldmath
\subsection*{$\cn =2$ minimal models}
\unboldmath

The $\cn =2$ minimal models correspond to the supersymmetric gauged WZW model
$SU(2)_k / U(1)$, and are characterized by the level $k$ of the supersymmetric
WZW model.
The $\cn =2$ minimal models characters are determined implicitly through the
identity:
\begin{equation}
\sum_{m \in \zi_{2k}} \mathcal{C}^{j\ (s)}_{m}  \Theta_{m,k} = \chi^{j}
\Theta_{s,2}\, ,
\end{equation}
where $\chi^j$ denotes a character of bosonic $SU(2)$ at level $k-2$. 
The characters are labeled by the triplet $(j,m,s)$. They correspond to the 
primaries of the coset $[SU(2)_{k-2}\times SO(2)_1]/U(1)_{k}$, however 
are not generically primaries of the $\cn = 2$ algebra. 
They have the R-charge
\begin{equation}
Q = \frac{s}{2}- \frac{m}{k} \mod 2
\end{equation}
The following identifications apply:
\begin{eqnarray*}
(j,m,s) &\sim &(j,m+2k,s)\\
(j,m,s) &\sim &(j,m,s+4)\\
(j,m,s) &\sim &(k/2-j-1,m+k,s+2)\\
\end{eqnarray*}
as well as the selection rule
\begin{equation}
2j+m+s =  0  \mod 2
\label{selruleMM}
\end{equation}
The weights of the primaries states are as follows:
\begin{equation}
\begin{array}{cccccc}
h &=& \frac{j(j+1)}{k} - \frac{n^2}{4k} + \frac{s^2}{8} \ & \text{for} & \ -2j \leqslant n-s \leqslant 2j \\
h &=& \frac{j(j+1)}{k} - \frac{n^2}{4k} + \frac{s^2}{8} + \frac{n-s-2j}{2}
\ & \text{for} & \ 2j \leqslant n-s \leqslant 2k-2j-4 \\
\end{array}
\end{equation}
We have the following modular S-matrix for these characters:\footnote{The reader may notice that the 
S-matrix of the N=2 minimal model given here may differ by a factor of two with the literature. Indeed in 
our conventions, the S-matrix is defined as acting on all triplets 
$(j,n,s)$, while it is often defined as acting on the fundamental domain w.r.t. the identification
$(j,n,s) \sim (k/2-j-1,n+k,s+2)$.
}
\begin{equation}
S^{j\, m\, s}_{\quad j' \, m' \, s'} = \frac{1}{2k} \sin \pi
\frac{(1+2j)(1+2j')}{k} \ e^{i\pi \frac{mm'}{k}}\ e^{-i\pi ss'/2}.
\end{equation}
Note also that the fusion rules of $SU(2)$ are given by:
\begin{equation}
N_{\hat{\jmath}\hat{\jmath}'}^j = 1 \ \text{for}\ |\hat{\jmath}-\hat{\jmath}'|\leqslant 
j \leqslant \text{min}\{\hat{\jmath}+\hat{\jmath}',k-\hat{\jmath}-\hat{\jmath}'\}\ \text{and}\ 
j+\hat{\jmath}+\hat{\jmath}' \in \zi \ , \quad 0 \ \text{otherwise.}
\end{equation}

\subsection*{Supersymmetric $SL(2,\mathbb{R})/U(1)$}
The characters of the $SL(2,\mathbb{R})/U(1)$ super-coset
at level $k$ come in
different categories corresponding to the classes of
irreducible representations
of the $SL(2,\mathbb{R})$ algebra in the parent theory. These coset characters 
coincide with the characters of the irreducible representations of the 
$\cn = 2$ superconformal algebra with $c>3$. In all cases
the quadratic Casimir of the representations is $c_2=-j(j-1)$.
First we consider \emph{continuous representations}, with $j = 1/2 + ip$,
$p \in \mathbb{R}^+$. The characters are denoted by
 $ch_c (p,m) \oao{a}{b}$,
where the $N=2$ superconformal
 $U(1)_R$ charge of the primary is $Q=2m/k$.
Then we have \emph{discrete representations} with $1/2 < j < (k+1)/2$,
of characters $ch_d (j,r) \oao{a}{b}$, where the $\cn=2$ $U(1)_R$
 charge of the primary is $Q=2(j+r+a/2)/k$,
$r\in \zi$. The third category corresponds to the \emph{finite
representations}, with $j=(u-1)/2$ and where $u = 1,2,3, \dots$
denotes the dimension of the finite representation.
These representations are not unitary except for the trivial representation with $u=1$.
The character for this identity representation
we denote by $ch_\mathbb{I} (r) \oao{a}{b}$. We can also 
define characters labeled by a $\zi_4$ valued quantum number for
$SL(2,\mathbb{R})/U(1)$, following the method we used to define
these characters for the free fermions. In other words, we define
these characters by summing over untwisted and twisted NS or R
sectors with the appropriate signs.

It is often convenient to define \emph{extended characters}, provided that $k$ is rational. In our 
particular example $k=2n/(n+2)$, and they are defined by summing over by summing
over $2n$ units of spectral flow.\footnote{If $n$ is even then one can reduce further 
the characters by summing over $n$ units of spectral flow.} 
These characters correspond to an extended chiral algebra which can be constructed along
the line of the extended chiral algebra for a $U(1)$ boson at rational
radius squared. For example, for
the continuous characters we define the corresponding extended characters
(denoted by capital letters) by:
\begin{equation}
  Ch_c (P,M)\oao{a}{b} = \sum_{w \in \zi} ch_c \left( P,\frac{M}{2(n+2)}+2nw \right) \oao{a}{b}
   = \frac{q^{\frac{(n+2)P^2}{2n}}}{\eta^3} \ \Theta_{M,2n(n+2)} \ \vartheta \oao{a}{b}.
\end{equation}
They carry a $\zi_{4n(n+2)}$ charge given by $M \in \zi$. In our conventions the 
$J_3$ eigenvalue of $\slr$ is $m=\nicefrac{M}{2(n+2)}$. In contrast with standard
characters, their modular transformations involve only a discrete set
of $N=2$ charges. 

Let us now consider briefly the discrete representations. They appear in the closed string spectrum
in the range $1/2 < j < (k+1)/2$. The primary states are:
\begin{eqnarray*}
|j,m=j+r\rangle&=&  |0\rangle_\textsc{ns} \otimes |j,m=j+r\rangle_\textsc{sl(2,r)} \quad r \geqslant 0 \\
|j,m=j+r\rangle &=& \psi^{-}_{-\frac{1}{2}}|0\rangle_\textsc{ns} \otimes (J^{-}_{-1} )^{-r-1}
|j,j\rangle_\textsc{sl(2,r)} \quad r<0 \\
\end{eqnarray*}
Thus in the $\zi_4$ formalism the former are in the $s=0$ sector and the latter in the $s=2$ sector.
These \textsc{ns} primary states have weights
\begin{eqnarray*}
h&=& \frac{j(2r+1)+r^2}{k}   \quad \ \ \qquad \qquad r \geqslant 0\\
h&=& \frac{j(2r+1)+r^2}{k} -r - \frac{1}{2} \qquad r<0\\
\end{eqnarray*}
To define conveniently the extended discrete characters, we adopt the following notation: 
\begin{equation}
Ch_d \oao{a}{b} (j,M) = \sum_{w \in \zi} ch_d \left( P,\frac{M}{2(n+2)}+2nw-j-\frac{a}{2} \right) \oao{a}{b}
\end{equation}
with the convention that this character is zero whenever $\frac{M}{2(n+2)}+2nw-j-\frac{a}{2}$ is not 
an integer. We don't need the explicit expression of these characters for our purposes; it can be found 
e.g. in~\cite{Israel:2004jt}

To compute the annulus amplitude between localized D-branes in the bulk of the text, 
we need the modular transformation of the extended character associated to the trivial
representation (in the $\zi_4$ formalism for fermions) that appears in the open 
string spectrum. It is given by~\cite{Eguchi:2003ik,Israel:2004jt}
\begin{eqnarray}
Ch_{\mathbb{I}}^{(s)} (r ;-1/\tau,0 ) \  = \sum_w ch_{\mathbb{I}}^{(s)} (r+2nw ;-1/\tau,0 )
  \hskip5cm
\nonumber\\
= \frac{1}{2n} \sum_{s' \in \zi_4} e^{-i \pi \frac{ss'}{2}}
\left\{
\int_{0}^{\infty} \di p' \!\!\!\!\! \sum_{M' \in \zi_{4n(n+2)}}\!\!\!\!\!\!\! e^{-i\pi \frac{r M'}{n}}
\frac{ \sinh 2\pi p'  \sinh [\pi (n+2) p' /n]}{\cosh 2\pi p' + \cos \pi \left(\frac{M'}{n+2}-s'\right)}
 Ch_{c}^{(s')} (p',M';\tau,0)  \right. \nonumber\\
\left.  +  \sum_{2(n+2)j'=n+3}^{3n+1}\  \sum_{r' \in \zi_n}
\sin \left(\frac{(n+2)\pi}{2n} (2j'-1) \right)\,  e^{-2i \pi \frac{(n+2)(j'+r')r}{n}}
\, Ch_{d}^{(s')} (j',r' , \tau ,0) \right\}\nonumber\\
\end{eqnarray}
The spectrum of primaries in the NS sector for this identity representation is as follows. First we have the identity operator $|r=0\rangle_\textsc{sl(2,r)} \otimes
| 0\rangle_\textsc{ns}$ belonging to the sector $s=0$.
The other primary states are:
\begin{eqnarray*}
|r\rangle&=& \psi^{+}_{-\frac{1}{2}} |0\rangle_\textsc{ns} \otimes (J^{+}_{-1} )^{r-1} |r=0\rangle_\textsc{sl(2,r)} \quad r>0 \\
|r\rangle &=& \psi^{-}_{-\frac{1}{2}}|0\rangle_\textsc{ns} \otimes (J^{-}_{-1} )^{-r-1} |r=0\rangle_\textsc{sl(2,r)}
\quad r<0 \\ \end{eqnarray*}
They belong to the sector $s=2$ and have weights
\begin{eqnarray*}
h&=& \frac{r^2}{k} +r - \frac{1}{2} \quad r>0\\
h&=& \frac{r^2}{k} -r - \frac{1}{2} \quad r<0\\
\end{eqnarray*}
the Ramond sector is obtained by one-half spectral flow. In particular, $r$ will be half-integer.

%%%%%%%%%%%%%%%%%%%%%%%%%%%%%%%%%%%%%%%%%%%%%%%%%%%%%%%%%%%%%%%%%%%%%%%%%%%%%%%%%%%%%%%%%%%%%%%%
%%%%%%%%%%%%%%%%%%%%%%%%BIBLIO%%%%%%%%%%%%%%%%%%%%%%%%%%%%%%%%%%%%%%%%%%%%%%%%%%%%%%%%%%%%%%%%%%
%%%%%%%%%%%%%%%%%%%%%%%%%%%%%%%%%%%%%%%%%%%%%%%%%%%%%%%%%%%%%%%%%%%%%%%%%%%%%%%%%%%%%%%%%%%%%%%%
%%%%%%%%%%%%%%%%%%%%%%%%%%%%%%%%%%%%%%%%%%%%%%%%%%%%%%%%%%%%%%%%%%%%%%%%%%%%%%%%%%%%%%%%%%%%%%%%

\end{document}